\documentstyle[emulateapj,graphicx]{article}

\newcommand{\temphi}{\left( \frac{T}{8000 \, {\rm K}} \right)}
\newcommand{\tempeight}{\left( \frac{T}{8000 \, {\rm K}} \right)}

\newcommand{\msolar}{{\rm M_{\odot}}}
\newcommand{\Htwo}{{\rm H_2}}
\def\gsim{\;\rlap{\lower 2.5pt 
\hbox{$\sim$}}\raise 1.5pt\hbox{$>$}\;}
\def\lsim{\;\rlap{\lower 2.5pt
   \hbox{$\sim$}}\raise 1.5pt\hbox{$<$}\;}
\newcommand{\redshift} {\left( \frac{1+z}{10} \right)}

\newcommand{\density} {\left( \frac{n_{\rm g}}{1 \, {\rm cm^{-3}}} \right)}

\newcommand{\denfour} {\left( \frac{n_{\rm g}}{10^{4} \, {\rm cm^{-3}}} \right)}

\newcommand{\beq}{\begin{equation}}
\newcommand{\eeq}{\end{equation}}
\def\myputfigure#1#2#3#4#5%
{\hskip0.03\textwidth\vskip#5pt
\makebox[0pt]{\hskip#2in
\includegraphics[width=#3\textwidth]{#1}}\vskip#4pt\hfill}


\lefthead{Oh \& Haiman}
\righthead{Collapse of Halos with $T_{\rm vir} > 10^4$K}

\begin{document}
\title{Second-Generation Objects in the Universe: Radiative Cooling and
Collapse of Halos with Virial Temperatures Above $10^4$ Kelvin}
\author{S. Peng Oh}
\vspace{0.1\baselineskip}
\affil{Theoretical Astrophysics, Mail Code 130-33, Caltech, Pasadena, CA,
91125, USA\\peng@tapir.caltech.edu}
\vspace{0.5\baselineskip}
\author{Zolt\'an Haiman\altaffilmark{1}} \affil{Princeton University
Observatory, Princeton, NJ 08544, USA\\ zoltan@astro.princeton.edu}
\altaffiltext{1}{Hubble Fellow}

\vspace{\baselineskip}
\begin{abstract}
The first generation of stars is thought to have formed in low-mass
halos with $T_{\rm vir} < 10^4$K where ${\rm H_{2}}$ cooling is
paramount. However, the efficiency of ${\rm H_{2}}$ formation and
cooling in these halos may have been severely limited by feedback processes. 
In this paper we investigate the radiative cooling and collapse of halos with
virial temperatures $T_{\rm vir} > 10^4$K, i.e. those that can cool in
the absence of ${\rm H_2}$ via neutral atomic lines.  The evolution of
these halos differs from their less massive counterparts.  Efficient
atomic line radiation allows rapid cooling to
$\sim 8000$ K; subsequently the gas can contract nearly isothermally
at this temperature.  In the absence of ${\rm H_2}$ molecules, the gas
would likely settle into a locally stable disk and only disks with
unusually low spin would be unstable.
However, we find that the initial atomic line cooling leaves a large,
out--of--equilibrium residual free electron fraction. This allows the
molecular fraction to build up to a universal value of $x_{\rm
H_2}\approx10^{-3}$, almost independently of initial density and
temperature. We show that this is a non--equilibrium freeze--out value 
that can be understood in terms of timescale arguments. Unlike in less massive halos,
$\Htwo$ formation and cooling is largely impervious to feedback from
external UV fields, due to the high initial densities achieved by
atomic cooling.  The newly formed molecules cool the gas further to
$\sim 100$ K, and allow the gas to fragment on scales of a few $\times
100$~M$_\odot$. We investigate the importance of various feedback 
effects such as $\Htwo$ photodissociation from internal UV fields and
radiation pressure due to Ly$\alpha$ photon trapping, which are likely
to regulate the efficiency of star formation.
\end{abstract}
\section{Introduction}
\label{sec:introduction}

The first generation of stars and/or quasars must have formed out of
gas with primordial composition dictated by Big Bang
nucleosynthesis. Only after the first generation of stars explode as
supernovae and achieve widespread metal pollution is metal line
cooling possible. Thus, the star formation efficiency and the initial
mass function (IMF) of Population III stars is likely to be set by the
physical regime in which metal--free cooling can take place. In recent
years, the cooling of metal--free gas in the first halos that are able
to cool within a Hubble time $t_{\rm cool} < t_{\rm H}$, has come
under intensive study (see Abel \& Haiman 2000 for a recent
review). Such halos have low virial temperatures, $T_{\rm vir} <
10^4$K; at these temperatures ${\rm H_{2}}$ is formed via gas--phase
processes such as ${\rm H + e^{-} \rightarrow H^{-} + \gamma}$,
followed by ${\rm H^{-} + H \rightarrow H_{2} + e^{-}}$. Detailed
numerical simulations have shown convergence toward a regime ${\rm T
\sim 200~{\rm K}\,}$, ${\rm n \sim 10^{4} \, {\rm cm}^{-3}}$, dictated
by the thermodynamic properties of ${\rm H_2}$ (Abel et al. 2000;
Bromm et al. 2001a), which allows gas fragmentation into clumps of
mass $10^{2}-10^{3} \, {\rm M_\odot}$.

However, as the first stars begin to shine, they emit
photo-dissociating radiation in the Lyman--Werner bands (LW;
11.2--13.6 eV) to which the universe is optically thin, and further
${\rm H_2}$ formation and cooling can be suppressed both by external
(Haiman, Rees \& Loeb 1997; Haiman, Abel \& Rees 2000; Ciardi et al
2000; Machacek et
al. 2001; Ricotti, Gnedin \& Shull 2001a, 2001b) and internal (Omukai \& Nishi 1999; Glover \& Brand 2000)
sources of UV radiation. This may be partially alleviated if X--rays
with energies $\gsim 1$ keV (whose mean free path in a uniform IGM exceeds the Hubble
length) can boost the free electron fraction, and thus the ${\rm H_2}$
formation rate in cooling clumps (Haiman, Rees \& Loeb 1996; Haiman,
Abel \& Rees 2000; Oh 2001); note that such effects are unlikely to be
important in the low density environment of the IGM (Venkatesan,
Giroux, \& Shull, 2001). Alternatively, protogalactic shocks can 
produce substantial ${\rm H_2}$ (Shapiro \& Kang 1987; Ferrara 1998),
or $\Htwo$ may be generated in front of HII regions or in relic HII
regions (Ricotti, Gnedin \& Shull 2001a, 2001b), or the optical depth
in intergalactic $H_{2}$ may reduce the strength of the external
dissociating radiation field (Ricotti, Gnedin \& Shull 2000). However, while these effects may
promote ${\rm H_2}$ formation in the presence of an external UV
background radiation field, they are unlikely to balance the strong
internal feedback provided by stars within a galaxy: in the worst case
scenario, only one star per halo can be formed (e.g., Madau
\& Rees 2001). It is therefore often believed that efficient and widespread 
star (and/or quasar black hole) formation capable of reionizing the
universe must await the collapse of
halos with $T_{\rm vir} > 10^4$K, or $M_{\rm halo} > 10^8
[(1+z)/11]^{-3/2} \, {\rm M_\odot}$ (Ostriker \& Gnedin 1996, Haiman,
Rees \& Loeb 1997, Ciardi et al 2000, Abel \& Haiman 2000, Ricotti,
Gnedin \& Shull 2001a, 2001b). These halos do not rely on the presence
of ${\rm H_2}$ molecules, since they can cool via recombination and
collisional excitation of neutral atoms. 
   
The goal of this paper is to critically examine the prevailing
assumptions of efficient gas cooling and star formation in metal free
halos with $T_{\rm vir} > 10^4$ K. To date, the details of gas cooling
and chemistry in these halos have not been studied with the same care
and attention devoted to lower mass halos.  Since the majority of
stars and/or quasars which reionized the universe, and polluted the
intergalactic medium (IGM) with metals, are expected to form in such
halos, it is important to study the evolution of the gas in these
halos in detail.

It is generally taken for granted that Ly$\alpha$ cooling of neutral
atomic hydrogen allows rapid contraction of gas until it becomes
self-gravitating at the center of the potential well and fragments to
form stars (ideas tracing back to Rees \& Ostriker 1977; White \& Rees
1978).  However, it is also known that significant contraction
is required for this fragmentation to result in stellar--mass
fragments.  If only Ly$\alpha$ cooling operates, the gas remains at a
temperature of $\sim 10^4$K due to the sharp cutoff in the
(equilibrium) cooling function and the Jeans mass is exceedingly high,
even at high densities: $M_{\rm J}\approx 10^{8} (T/10^4~{\rm
K})^{3/2} (n/1~{\rm cm^{-3}})^{-1/2} \, \msolar$. Even if the gas
became self-gravitating, unless the gas can contract to extremely high
densities, $n > 10^{12} {\rm cm^{-3}}$, fragmentation to lower Jeans masses
$M \sim 100 {\rm M_\odot}$ cannot proceed. However, the gas cannot
cool to arbitrarily high densities, but eventually must form a
rotationally supported disk; at the temperatures allowed by Ly$\alpha$
cooling, we shall show that the majority of such disks are locally
gravitationally stable.

Thus, an additional coolant is needed to lower the gas temperature,
both to ensure gravitational instability and to lower the Jeans mass
by several orders of magnitude. In the absence of metals, ${\rm H_2}$
formation and cooling is therefore still critical to star formation in
$T_{\rm vir} > 10^4$K halos, and cannot be ignored.  In this paper, we
study ${\rm H_2}$ formation in shock--heated gas, and show that a
universal ${\rm H_2}$ fraction of $x_{\rm H_2} \approx 10^{-3}$ forms
in gas that cools from an initial temperature of $T > 10^4$K. Similar behavior has already been noted in previous studies of
pregalactic shocks (Shapiro \& Kang 1987, Kang \& Shapiro 1992). Here
we study the non-equilibrium ${\rm H_2}$ formation in the $T_{\rm vir}
> 10^4$K halos of interest, examine the robustness of this mechanism
to variations in density, temperature, and radiation field, and show
that the asymptotic abundance can be understood in terms of timescale
arguments.  These arguments reveal that over a wide range of
densities, the cooling gas follows a universal track in the $(x_{\rm
e},T)$ plane.

This paper is organized as follows. In \S\ref{disk}, we study the
equilibrium structure of isothermal disks embedded in dark matter
halos, and show that ${\rm H_2}$ cooling is needed to promote
gravitational instability in most disks. In \S\ref{cool}, we use
semi--analytic methods and non--equilibrium chemistry to investigate
${\rm H_2}$ formation and radiative cooling in halos with virial
temperatures $T_{\rm vir} > 10^4$K, and argue that the ${\rm H_2}$
abundance builds up to a universal value of $x_{\rm H_2}\approx
10^{-3}$ under most realistic conditions. In
\S\ref{feedback}, we study the effects of ${\rm H_{2}}$ destruction by
internal and external sources of UV radiation and show that feedback
processes are much less efficient in $T_{\rm vir} > 10^{4}$K halos than in
their smaller counterparts, primarily because in the larger halos, gas
can be compressed to high densities by initial atomic line cooling. We
also examine whether opacity and radiation pressure effects can halt
the collapse or fragmentation. Although we conclude that this is
unlikely, we argue that it could affect the efficiency of star
formation.  In
\S\ref{conclusions}, we summarize our conclusions and
the implications of this work. In all numerical estimates, we assume a
cold dark matter cosmology with a cosmological constant ($\Lambda$CDM)
$(\Omega_{\rm m},\Omega_{\Lambda},\Omega_{\rm b}h^{2},h,\sigma_{8
h^{-1}})=(0.3,0.7,0.019,0.7,1.0)$ (see, e.g., Bahcall et al. 1999 for
a review of these choices).

\section{Disk Formation and Gravitational Instability}
\label{disk}

Let us first consider the cooling and collapse of an initially
spherical configuration of gas in a typical halo with $T_{\rm vir} >
10^4$K. The discussion presented in this section serves two purposes:
it will highlight the importance of ${\rm H_2}$ molecules for the
larger halos, and will also yield the physically appropriate range of
density, temperature and ionization fractions under which to consider
${\rm H_2}$ formation and cooling in later sections.

The well--known condition for runaway contraction of gas is $t_{\rm
cool} < t_{\rm dyn} < t_{\rm sc}$ (see Rees \& Ostriker 1977), which
is typically satisfied for hydrogen atomic line cooling. Since cooling due to Ly$\alpha$ has a
very sharp cutoff at $T < 10^4$K, the cooling time is a very sensitive function of
temperature, increasing rapidly by several orders of magnitude in a
very narrow temperature range below $\sim 10^{4}$K (Spitzer
1978). This is because the free electron fraction able to excite
atomic line cooling drops very rapidly in this temperature range. Thus, the condition $t_{\rm cool} < t_{\rm dyn}$ implies
that the collapse is nearly isothermal. If the gas cools below $\sim
10^4$K, the gas recombines, the cooling time rapidly increases until
$t_{\rm cool} > t_{\rm dyn}$, and further contraction of the gas is
close to adiabatic, with the gas heating up due to the contraction
until the hydrogen atoms are collisionally re--ionized and the
condition $t_{\rm cool} < t_{\rm dyn}$ is satisfied once again. In the
absence of any other effects, this thermostatic mechanism would allow
the gas in a $T_{\rm vir} > 10^{4}$K halo to cool and contract to
arbitrarily high densities.  In practice, however, the gas has some
initial angular momentum, and must therefore eventually become
rotationally supported. After a contraction by a factor of
$\lambda^{-1}\sim 20$ in radius (where $\lambda \equiv J |E|^{1/2}/G
M^{5/2}$ is the spin--parameter, and $J$, $E$, and $M$ are the total
angular momentum, energy and mass of the halo), this results in rotationally supported disk at the
center of the halo (Mo, Mao \& White 1998, hereafter MMW). It is
possible that some fragmentation takes place as the gas collapses
toward a disk, however, this is likely to be inefficient due to the slow growth of density fluctuations in a rapidly contracting medium (e.g.,
Kashlinsky \& Rees 1983). We therefore assume that most fragmentation
must take place in the disk itself. We shall show that if only atomic
line cooling operates, the disk will be stable to fragmentation in the
majority of cases. For fragmentation to proceed, an additional cooling
channel (such as $H_{2}$) is required to allow cooling below $\sim 10^{4}$K. 

For simplicity, we shall follow MMW, and assume that the gas settles
to an isothermal, exponential disk, with gas temperature $T_{\rm
gas}$, embedded in a halo of virial temperature $T_{\rm vir}$ with a
Navarro, Frenk \& White (1997, hereafter NFW) dark matter density
profile.  We begin by listing the characteristic properties of such
disks, which will be relevant for our later studies of ${\rm H_2}$
formation and self--shielding. Let us assume that baryons make up the
universal mass fraction $\Omega_{\rm b}/\Omega_{\rm m}$ of the halo,
of which some fraction $f_{\rm d}$ have collapsed into the disk,
i.e. $M_{\rm disk}=m_{\rm d} M_{\rm halo}= f_{\rm d} (\Omega_{\rm
b}/\Omega_{\rm m}) M_{\rm halo}$. We also assume that the angular
momentum of the disk $J_{\rm d}$ is some fraction $j_{\rm d}$ of the
halo angular momentum $J$, i.e. $J_{\rm d}=j_{\rm d} J$. Henceforth,
we shall assume that the specific angular momentum of the disk is
similar to that of the halo, and thus $m_{\rm d}=j_{\rm d}$.

The assumption that the baryons preserve their specific angular
momentum during collapse results in a good fit to the observed size
distribution of galactic disks (Mo, Mao \& White 1998). Detailed
numerical simulations (Navarro \& White 1993, Navarro \& Steinmetz
2000) have not supported this simple model, but instead produced
significantly smaller disks, due to the transfer of angular momentum
from the gas to the dark matter during the highly inhomogeneous
collapse.  These simulations, however, also fail to produce the sizes
and properties of observed galaxy--sized disks. Inclusion of
suppression of cooling until late times (Weil, Eke \& Efstathiou 1998)
or supernovae feedback (Thacker \& Couchman 2001) reduces the
discrepancy, but the issue has yet to be conclusively resolved. We
therefore simply extend successful semi-analytic models of disk
formation at low redshift to higher redshift, and note that future
numerical simulations of such halos may not in fact produce such
disks.

The hydrogen number density at radius $r$ and at vertical height $z$
in an isothermal exponential disk of radial scale length $R_{\rm d}$
is given by (Spitzer 1942)
\begin{equation}
n(r,z)=
 n_{o} 
 {\rm exp}\left(-\frac{2 r}{R_{\rm d}} \right) 
 {\rm sech^{2}} \left(\frac{z}{\sqrt{2} z_{o}} \right),
\label{disc_density}
\end{equation}
where $n_o$ is the central density, $z_o$ is the vertical scale height
of the disk at radius $r$,
\begin{equation}
 z_{o}= \frac{c_{\rm s}}{(4 \pi G \mu m_{\rm H} n_{o} e^{-2r/R_{\rm d}})^{1/2}}
\end{equation}
$c_{\rm s}$ is the sound speed of the gas, and $\mu=0.6$ is the mean
molecular weight (we choose the definition of the disk scale length
$R_{d}$ to conform to the customary assumption that the surface
density has an exponential profile: $\Sigma \propto {\rm
exp}(-r/R_{d})$). For a disk in a halo with spin parameter $\lambda
$, if we assume the baryons
conserve their specific angular momentum when they collapse, then the disk
scale length is given by $R_{\rm d} = 2^{-1/2} (j_{\rm d}/m_{\rm d})
\lambda r_{200} f_{c}^{-1/2} f_{R} \approx \frac{\lambda}{\surd 2} r_{200}$,
where $r_{200} \approx r_{\rm vir}$ is the radius that encloses a mean
interior mass density of 200$\rho_{\rm crit}$, $f_{c}(c)$ and
$f_{R}(\lambda,c,m_{\rm d},j_{\rm d})$ are dimensionless functions of
order unity (MMW), and $c$ is the dimensionless concentration
parameter (NFW).

The central number density of the gas is obtained by setting $\int dz
\int 2 \pi r \, dr \mu m_{\rm p} n(r,z) = M_{\rm disk}$, which yields:
\begin{eqnarray}
\nonumber
n_{o} &\approx& 3.2 \times 10^{4} 
\left( \frac{f_{\rm d}}{0.5} \right)^{2} 
\left( \frac{T_{\rm gas}}{8000 \, {\rm K}} \right)^{-1} 
\left( \frac{T_{\rm vir}}{5 \times 10^{4} \, {\rm K}} \right) \\
&&
\times
\left( \frac{\lambda}{0.05} \right)^{-4} 
\left( \frac{1+z}{10} \right)^{3} \ {\rm cm^{-3}}
\label{central_density}
\end{eqnarray}
When considering the characteristic densities for ${\rm H_2}$
formation, it will be sufficient to consider densities at most an
order of magnitude below the central density: gas with $n > 0.1 n_{o}$
comprises $> 50 \%$ of the mass of the disk.  Because hydrostatic
support is only relevant in the vertical direction, as the gas cools
the disk becomes thinner, with a reduced scale height $z_{o} \propto
T_{\rm gas}^{1/2}$. The vertical column density of gas as a function
of radius is:
\begin{eqnarray}
\nonumber
N_{\rm HI}(r) &=& \sqrt{2} n_{o} z_{o} {\rm exp}(-2 r/R_{\rm d})  \\ 
\nonumber
 &=& 
3 \times 10^{23}\, {\rm exp}(-r/R_{\rm d}) 
\left( \frac{T_{\rm vir}}{5 \times 10^{4} \, {\rm K}} \right)^{1/2} 
\times \\
 &&
\left( \frac{f_{\rm d}}{0.5} \right) 
\left( \frac{\lambda}{0.05} \right)^{-2} 
\left( \frac{1+z}{10} \right)^{3/2} \ {\rm cm^{-2}}
\label{column_density}
\end{eqnarray}
which is sufficient for self-shielding of the gas against both
ionizing UV radiation and H2 dissociating radiation in the LW bands to
become important. Note that the column density is independent of the
gas temperature.

We now consider the conditions for gravitational instability of the
disk. In computing the rotation curve $V(r)$, we use the formalism of
MMW, which takes into account the contraction induced in the inner
regions of the halo by the cooling and formation of the disk. This is
done by assuming the disk is assembled slowly and the angular momentum
of dark matter particles is an adiabatic invariant (Blumenthal et
al. 1986, Flores et al. 1993). For the disk to be locally
gravitationally unstable despite the stabilizing effects of tidal
shears and pressure forces, we require the Toomre parameter $Q <1$,
where (e.g. Binney \& Tremaine 1987)
\begin{equation}
Q= \frac{c_{\rm s} \kappa}{\pi G \Sigma},
\label{toomre}
\end{equation}
$\Sigma$ is the disk surface mass density, and $\kappa = 1.41 (V/r) (1
+ {\rm d \, ln} V/{\rm d \, ln} r)^{1/2}$ is the epicyclic
frequency. Regions in local disk galaxies where $Q>1$ are
observationally associated with very little star formation, indicating
that the Toomre criterion is obeyed remarkably well (Kennicutt
1989). For our purposes, if $Q > 1$ everywhere throughout the disk, it
is gravitationally stable and we assume no star formation takes
place. Disks with high spin parameters have low surface densities and
satisfy this criterion. For any given disk--halo system, we can
calculate a critical spin parameter $\lambda_{\rm crit}$ for which the
disk is marginally stable. Here we define $\lambda_{\rm crit}$ by the
requirement that $Q$ attains a minimum value of $Q=1$ at least at one
position in the disk. The fraction of halos that remain dark can
therefore be found by integrating over the spin parameter
distribution, $f_{\rm dark} = \int_{\lambda_{\rm crit}}^{\infty}
p(\lambda) d\lambda$, where $p(\lambda)$ is given by
\begin{equation}
p(\lambda) d\lambda = \frac{1}{\sigma_{\lambda} (2 \pi)^{1/2}} {\rm
exp} \left( - \frac{{\rm ln}^{2} (\lambda/\bar{\lambda})}{2
\sigma_{\lambda}^{2}} \right) \frac{d\lambda}{\lambda}
\label{spin_dist}
\end{equation}
with $\bar{\lambda}= 0.05$ and $\sigma_{\lambda} =0.5$ (e.g. Barnes \&
Efstathiou 1987, Warren et al. 1992).\footnote{In another context, it
was proposed that similarly high spin dark halos might be detected
from their gravitational lensing signature (Jimenez et al. 1997).} Our
use of the Toomre criterion to characterize gravitational instability
depends on the assumption that the disk is thin, $z_{o} \ll R_{\rm
d}$; in this case gravitational instability is insensitive to the
vertical structure of the disk (Goldreich \& Lynden-Bell 1965). The
ratio of scale heights is given by
\begin{equation}
\frac{z_{o}}{R_{\rm d}} = 
1.5 \times 10^{-2} 
\left( \frac{T_{\rm gas}}{8000 \,{\rm K}} \right) 
\left( \frac{T_{\rm vir}}{5 \times 10^{4}\, {\rm K}} \right)^{-1} 
\left( \frac{\lambda}{0.05} \right)^{-1} 
\left( \frac{f_{\rm d}}{0.5}\right)
\end{equation}
and thus $z_{o} \ll R_{\rm d}$ is generally satisfied, except in low
virial temperature halos where $T_{\rm vir} \sim T_{\rm gas}$.

\myputfigure{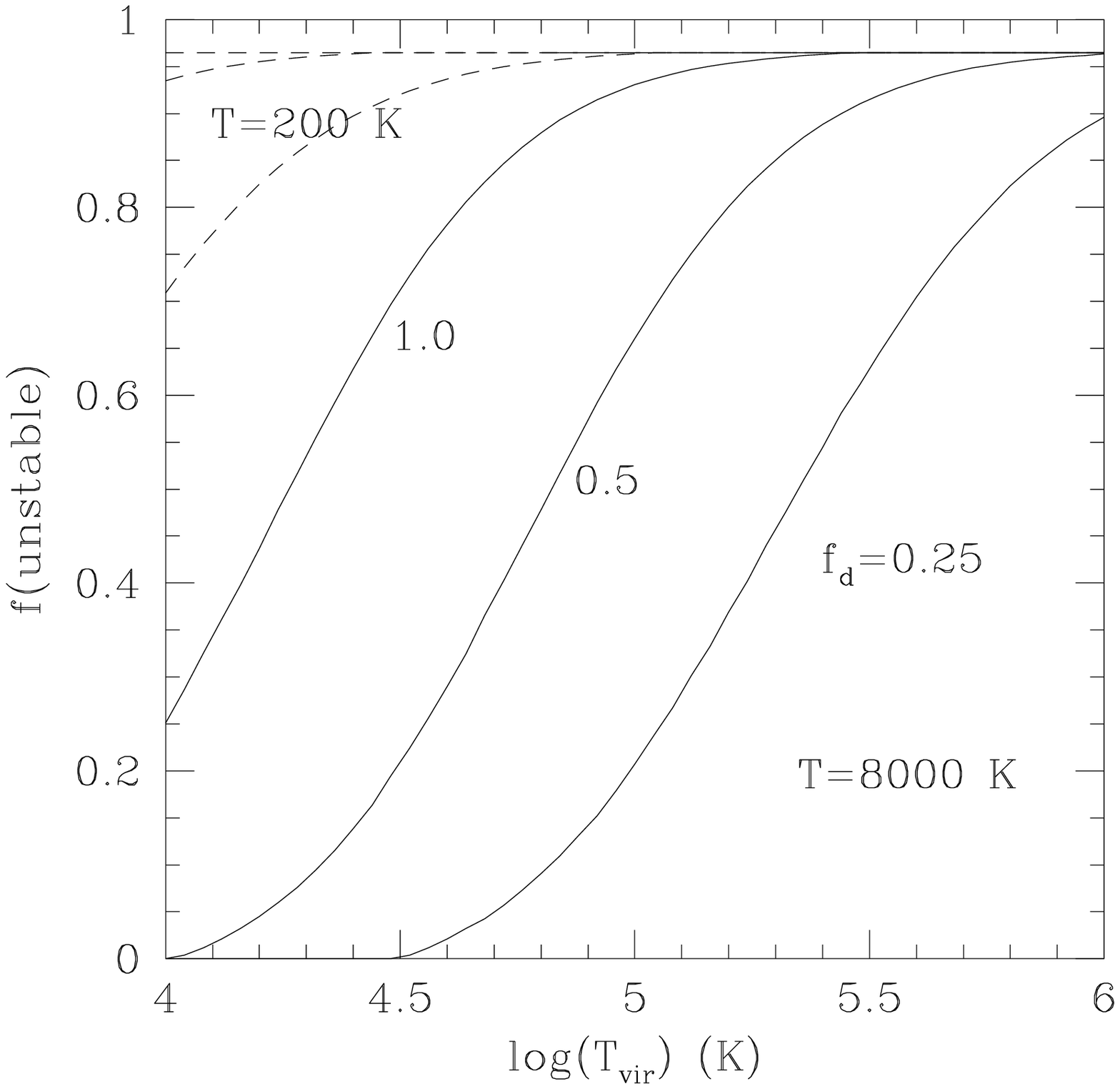}{3.3}{0.5}{-25}{-35}
\figcaption{\label{frac_spiral}
The fraction of disks that have sufficiently low spin to be locally
gravitationally unstable at least at one radius, as a function of
virial temperature $T_{\rm vir}$ of the halo. The computed curves are
for z=10, but the results depend only very weakly on redshift. The solid and dashed
curves assume gas temperatures of 8000 K (no $\Htwo$) and 200 K
(effective $\Htwo$ cooling), respectively. For
both temperatures, results are shown for three different disk
mass--fractions, $f_{\rm d}\equiv (M_{\rm disk}/M_{\rm
halo})/(\Omega_{\rm b}/\Omega_{\rm m})=0.25, 0.5$, and 1. For $f_{\rm
d}=0.25$, (as inferred for low redshift dwarfs), the majority of disks
in low $T_{\rm vir}$ halos will be gravitationally stable.}
\vspace{\baselineskip}

>From equation (\ref{toomre}), it is clear that if the gas cools from
$\sim 8000$K to $\sim 200$K, this lowers by a factor of $\sim 6$ the
critical surface density required for local gravitational
instability. In Figure \ref{frac_spiral}, we show the fraction of disk
galaxies in halos with virial temperature $T_{\rm vir}$ that are
gravitationally unstable at redshift $z=10$, for different values of
$f_{\rm d}$. The most important result seen in this figure is that in
the absence of ${\rm H_2}$, i.e., at a temperature of $8000$K,
virtually all halos just above the threshold $T_{\rm vir}\approx
10^4$K harbor disks that are stable everywhere.  Instability could be
promoted if a large fraction, $f_{\rm d} > 0.5$, of the baryons make
up the disk, however, this is unlikely to be the case (see discussion
below). Only in relatively large halos ($T_{\rm vir}\gsim 10^5$K)
would most disks be able to develop instabilities. This can be easily
understood with a simple order of magnitude calculation assuming an
exponential disk which reaches the asymptotic circular velocity on a
disk scale length $R_{\rm d} \approx \lambda r_{200}/\surd 2$, which
gives: $Q \approx 1.3 (T_{\rm gas}/8000 {\rm K})^{1/2} (T_{\rm vir}/2
\times 10^{4} {\rm K})^{-1/2}$ (a more detailed calculation gives a
slightly lower value). Apart from a weak dependence on the
redshift-dependent NFW concentration parameter $c$ (at $z\sim 10$, we
have $c \approx 5$ for halos with $T_{\rm vir} \approx {\rm few}
\times 10^{4}$K), the Toomre parameter Q is independent of redshift
for a halo of given $T_{\rm vir}$ and $f_{\rm d}$, and thus the
results of Figure \ref{frac_spiral} depend very weakly on the assumed
redshift.

The figure also shows that decreasing the gas temperature
significantly increases the fraction of unstable disks, particularly
in low mass halos. To illustrate the effect of the gas temperature
more clearly, we have computed the mass--weighted fraction of disks,
embedded in all halos with $T_{\rm vir} > 10^{4}$K, that are unstable
to star formation:
\begin{equation}
\tilde{f}_{\rm unst}(z) = 
\frac{\int_{M(T_{\rm vir}=10^{4}{\rm K},z)}^{\infty} dM
\frac{dN}{dM}(M,z) M f_{\rm unst}(M,z)}{\int_{M(T_{\rm vir}=10^{4}{\rm K},z)}^{\infty} dM
\frac{dN}{dM}(M,z) M }
\label{frac_unstable_redshift}
\end{equation}
where the mass function $dN/dM(M,z)$ is obtained from standard
Press--Schechter theory, and $f_{\rm unst}(M,z)$ is the quantity
computed in Figure \ref{frac_spiral}.  Thus, $\tilde{f}_{\rm unst}$ is
the quantity directly relevant to most cosmogonic studies,
representing the fraction of the total gas mass reservoir that will be
available for star--formation.  The resulting unstable fraction
$\tilde{f}_{\rm unst}$ is shown as a function of redshift in
Figure~\ref{redshift_evol}.

The figure reveals that depending on the value of $f_{\rm d}$, the
temperature drop caused by $\Htwo$ formation and cooling can
significantly increase (by between one or two orders of magnitude) the
mass fraction of collapsed halos that are able to form stars. If
$\Htwo$ cooling does not proceed efficiently, then significant star
formation, and therefore the epoch of reionization, has to likely
await lower redshifts when still more massive ($T_{\rm vir}\gg10^4$K)
halos collapse, or until the dispersal of metals from low--spin,
star--forming halos sufficiently contaminates surrounding halos so
that they can cool via metal lines.

Since $\Sigma \propto M_{\rm disk}$, the question of whether a disk
will be gravitationally unstable depends strongly on the mass fraction
of baryons in the disk.  As clearly revealed in Figures
\ref{frac_spiral} and \ref{redshift_evol}, if $f_{\rm d}$ is high, a
larger fraction of disks will be gravitationally unstable. For
simplicity, here we have assumed $f_{\rm d}$ to be a constant, but in
reality it is likely to vary as a function of virial temperature.  One
might indeed expect that $f_{\rm d}$ is lower in halos with lower
virial temperatures, owing to the increased efficiency of feedback
processes in shallow potential wells. This would accentuate the trend
for disks in low mass halos to be gravitationally stable.  For
reference, $f_{\rm d} \approx 0.4$ yields good agreement with the
observed sizes of disks and rotation curves over a wide range of
circular velocities (MMW).  More appropriately for our purposes, van
den Bosch, Burkert \& Swaters (2001) fitted angular momentum models to
a sample of observed low--mass disk galaxies with ${\rm V_{\rm circ}
\sim 75-150 \, km \, s^{-1}}$ (still somewhat more massive than the
galaxies we are considering); they found best--fit mass fractions to
be considerably lower than the universal baryonic fraction, with a
mean value of $f_{\rm d} \sim 30\%$. Interestingly, they find that
although disks form out of only a small fraction of the available
baryons, they nonetheless draw most of the available angular
momentum. If disks formed only out of low angular momentum baryons in
the halo, they would be exceedingly compact, obviating our
conclusions.

\myputfigure{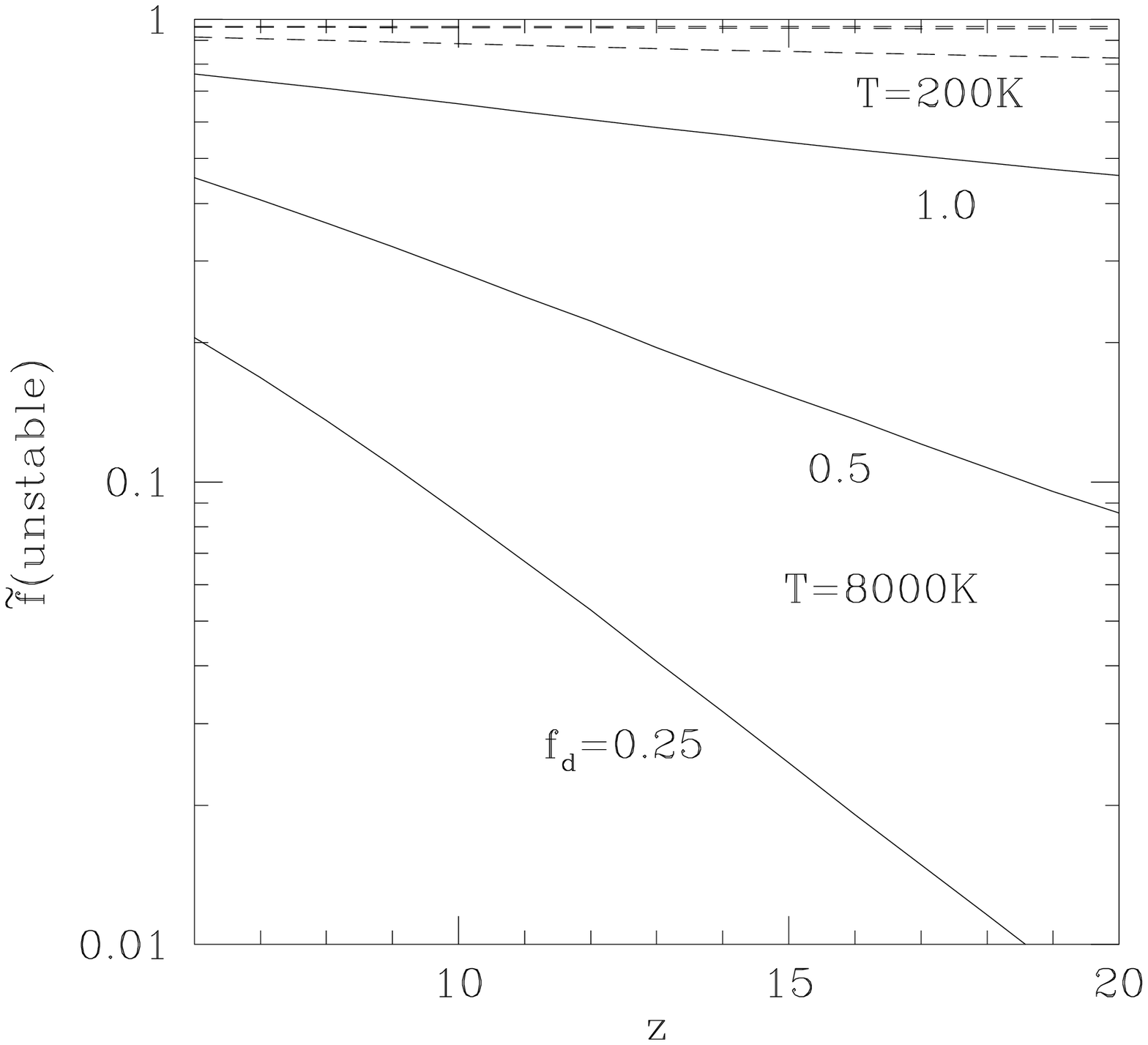}{3.3}{0.5}{-25}{-35}
\figcaption{\label{redshift_evol}
Mass--weighted unstable disk fraction as a function of redshift,
as given by equation (\ref{frac_unstable_redshift}). At every
redshift, the mass fraction in Toomre unstable disks have been summed over all halos with $T_{\rm
vir}\geq10^4$K. Thus, $\tilde{f}_{\rm unst}$ represents the fraction
of the total gas mass reservoir that will be available for
star--formation in all collapsed halos.  The curves are labeled as in
Figure~\ref{frac_spiral}. If $\Htwo$ cooling is ineffective, then the
gas fraction available for star formation in collapsed halos is considerably lower.} 
\vspace{\baselineskip}

It is not clear what direction $f_{\rm d}$ should take towards higher
redshift.  While Ly$\alpha$ cooling is likely to be more effective at
the higher densities, typical masses and virial temperatures are even
lower than those considered by van den Bosch et al. (2001), metal line
cooling is absent, and a possibly top--heavy IMF results in more
intense ionizing radiation (Bromm, Kudritzki \& Loeb 2001) once
internal star formation is underway and is also more likely to produce
hypernovae explosions (Heger \& Woosley 2001). All of these effects should
suppress gas cooling efficiencies and lower $f_{\rm d}$. If $f_{\rm d}\sim 30\%$, as
at low redshift, then from Figure \ref{redshift_evol}, only
$\sim$few$\%$ of the gas in disks will be available for star formation
at redshifts $z>10$.

In summary, our results in this section underscore the need for ${\rm
H_2}$ formation in order for most disks to be unstable. In the next
section, we will show that a significant amount of ${\rm H_2}$ is
indeed expected to form in virtually all realistic cases; while in
section \S\ref{feedback} we consider the feedback processes which
might suppress $\Htwo$ formation and cooling.

\section{Chemistry and Gas Cooling in $T_{\rm vir} > 10^4$K halos}
\label{cool}

In this section, we follow the coupled chemical and thermal evolution
of a single fluid element when no UV flux is present. We find that in
gas cooling from above $\sim 10^{4}$K, $\Htwo$ forms with a fixed
abundance $x_{\Htwo} \sim 10^{-3}$ , which is {\it not} its equilibrium value, under a wide variety of initial conditions. We obtain a quantitative explanation for this value and the reason for the universality of this abundance. Readers not interested in the technical details may skip
the rest of this section. We defer discussion of the case where UV
flux is present to \S 4.  

It is not immediately obvious that ${\rm H_2}$ can form in gas cooling
down from $T_{\rm vir} > 10^{4}$K, since destruction of ${\rm H_2}$ by
charge exchange and collisional dissociation is very effective in gas
at temperatures $T > 5000$K. There is a trough in the equilibrium
cooling curve between $\sim 8000$K (when Ly$\alpha$ cooling is
effective) and $\sim 3000$K (when ${\rm H_2}$ cooling becomes
effective), and it is possible that gas could cool to $\sim 5000$K and
recombine before sufficient $\Htwo$ forms to cool the gas to lower
temperatures, hanging up in the valley between the peaks of the two
cooling curves (e.g., see figure 12 of Barkana \& Loeb 2001). 

The chemistry and cooling of primordial gas cooling
from above $10^{4}$K has previously been considered by integrating the
coupled rate equations in the limit of steady state shock waves
(Shapiro \& Kang 1987, Kang \& Shapiro 1992, hereafter KS).  These
authors find that ${\rm H_2}$ formation is indeed possible and that
gas can cool continuously to $\sim 200$K, the limiting temperature at
which ${\rm H_2}$ molecules comes into local thermodynamic equilibrium
(LTE), and cooling is no longer effective. The efficacy of ${\rm H_2}$
formation is due to the large non-equilibrium abundance of electrons
in gas cooling from above $T > 10^{4}$K; the gas cools and forms ${\rm
H_2}$ faster than it recombines. Shapiro \& Kang (1987) and subsequent
authors have noted that ${\rm H_2}$ tends to form with an asymptotic
abundance of $\sim 10^{-3}$ over a wide range of initial
conditions. We confirm this result, explore the process of $\Htwo$
formation semi-analytically, and demonstrate how this asymptotic
abundance can be understood in terms of timescale arguments. Similar
results have been obtained recently by Susa et al. (1998).

\myputfigure{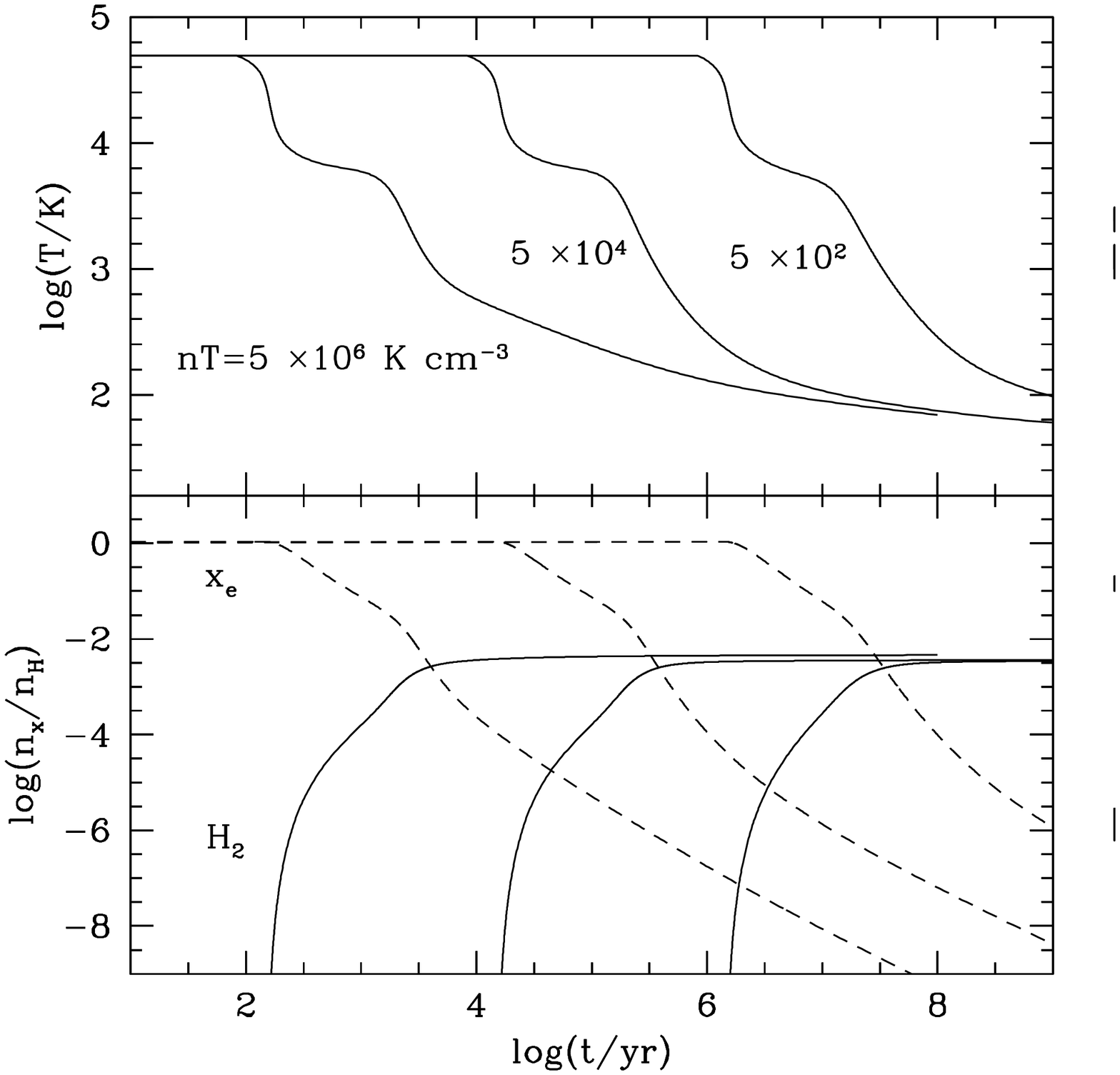}{3.3}{0.5}{-15}{-10}
\figcaption{\label{full_chem}
The temperature (top panel), $\Htwo$ and electron abundance (bottom
panel) evolution as a
function of time for a parcel of gas cooling isobarically from $T=5
\times 10^{4}$K, with no dissociating UV flux. Regardless of the
density of the gas, the $\Htwo$ abundance asymptotes to a universal
fraction $x_{\Htwo} \sim 10^{-3}$. The evolution of the gas is
self-similar, with timescales scaling as $1/n$, except at high gas
densities $n > 10^{4} {\rm cm^{-3}}$, when LTE effects
become important. Thus, for the nT=$5
\times 10^{6} {\rm K cm^{-3}}$ curve, the temperature evolution
deviates from self-similarity at low temperatures (high densities). See text for details.}
\vspace{\baselineskip}

\subsection{A universal ${\rm H_{2}}$ abundance}

Here, we show that over a wide range of initial conditions, $H_{2}$
forms in cooling gas with a universal abundance $x_{\rm H_{2}} \sim
10^{-3}$ which is {\it not} the equilibrium value. 

We begin by using a full non-equilibrium chemistry code to solve the rate equations for gas cooling from above $10^{4}$ K. The code solves the
coupled set of stiff equations using the Livermore stiff solver
LSODAR, assuming primordial abundances and starting from the initial
conditions $T_{\rm gas}=T_{\rm vir}$ and assuming equilibrium
ionization fractions at $T_{\rm vir}$ as the initial condition (for
$T_{\rm vir}= 5 \times 10^{4}$K, $x_{\rm e}=1$). The reaction network
we use in our calculations is given in a Table in the Appendix. The
rates are identical to those used in Haiman, Rees \& Loeb (1996),
unless marked otherwise. In Figure \ref{full_chem}, we show results from the code for gas cooling isobarically (the
results for isochoric cooling are very similar). For
a wide range of initial densities, the ${\rm H_2}$ abundance
asymptotes to a universal value $x_{\rm H_{2}} \sim 10^{-3}$. We also
obtain this universal ${\rm H_{2}}$ abundance for any starting temperature
$T_{vir} > 1.2 \times 10^{4}$K.  

Does this universal abundance simply reflect the equilibrium abundance
of ${\rm H_2}$ at low temperatures? The equilibrium abundance of $n_{\rm
H^{-}}$ is given by:
\begin{eqnarray}
\nonumber
n_{\rm H_{-}} =&& \frac{k_{9}n_{\rm H}n_{\rm e} + k_{14} n_{\rm H_{2}}n_{\rm e}} {(k_{10}+k_{20})
n_{\rm H} + (k_{13}+k_{21}) n_{\rm H^{+}} + k_{19} n_{\rm e}}   \\ 
\approx&& \frac{n_{\rm H}
n_{e^{-}} k_{9}}{n_{\rm H} k_{10} + n_{\rm H^{+}} k_{13}}
\approx \frac{n_{e^{-}} k_{9}}{k_{10} }
\label{Hminus}
\end{eqnarray}
The various approximations indicate the most important terms, which we
have identified directly from the non-equilibrium chemistry code. The
last approximation holds for $x_{\rm e} < 0.03$, when the $H_{2}$
abundance rises rapidly and approaches its asymptotic value. The equilibrium abundance of ${\rm H_2}$ is then given by:
\begin{eqnarray}
\label{H2_equil}
\nonumber
n_{\rm H_{2}} =&& \frac{k_{10} n_{\rm H} n_{\rm H^{-}}}{(k_{14}+k_{17}+k_{18}) n_{\rm H^{+}}
+k_{15} n_{\rm H}} \\  \approx&& \frac{k_{10} n_{\rm H}
n_{\rm H^{-}}}{k_{17} n_{\rm H^{+}}} \approx \frac{k_{9}n_{\rm H}}{k_{17}}
\end{eqnarray}
where again the last approximation holds for $x_{\rm e} < 0.03$. Note that in this regime $x_{\rm H_{2}} \approx
k_{9}/k_{17}$ is independent of density or ionization fraction and
depends only on the temperature. It is interesting to note that
collisional dissociation of ${\rm H_2}$, given by ${\rm H_{2} + H
\rightarrow 3H}$, is unimportant in this regime, contrary to previous
assumptions (e.g. Omukai 2001). Instead, the destruction of ${\rm
H_2}$ is governed primarily by the charge-exchange reaction, ${\rm
H_{2} + H^{+} \rightarrow H_{2}^{+} + H}$; when the gas is fully
ionized, ${\rm H_2}$ formation is strongly suppressed. This is why
recombination and ${\rm H_2}$ formation both initially proceed on the
same timescale in Figure \ref{full_chem}.

We now use these expressions to check if $\Htwo$ remains in
equilibrium throughout the cooling process. In the lower panel of Figure \ref{timescales}, we plot the equilibrium
abundance of ${\rm H_{2}}$ (as given by the full, rather than reduced
expressions for equations \ref{Hminus} and
\ref{H2_equil}), against the actual computed value of the $\Htwo$
abundance given by the chemistry code. Initially, the
${\rm H_{2}}$ abundance follows the equilibrium value. However, as the
gas cools to lower temperatures $T < 3700$K, it deviates sharply from
equilibrium: the ${\rm H_{2}}$ abundance asymptotes to the value
$x_{\rm H_2}^{\rm asymp} \sim 10^{-3}$, while the equilibrium value continues to rise. Thus, the universal abundance does {\it not} reflect the equilibrium abundance of
${\rm H_2}$ at low temperatures, which is significantly higher.

\myputfigure{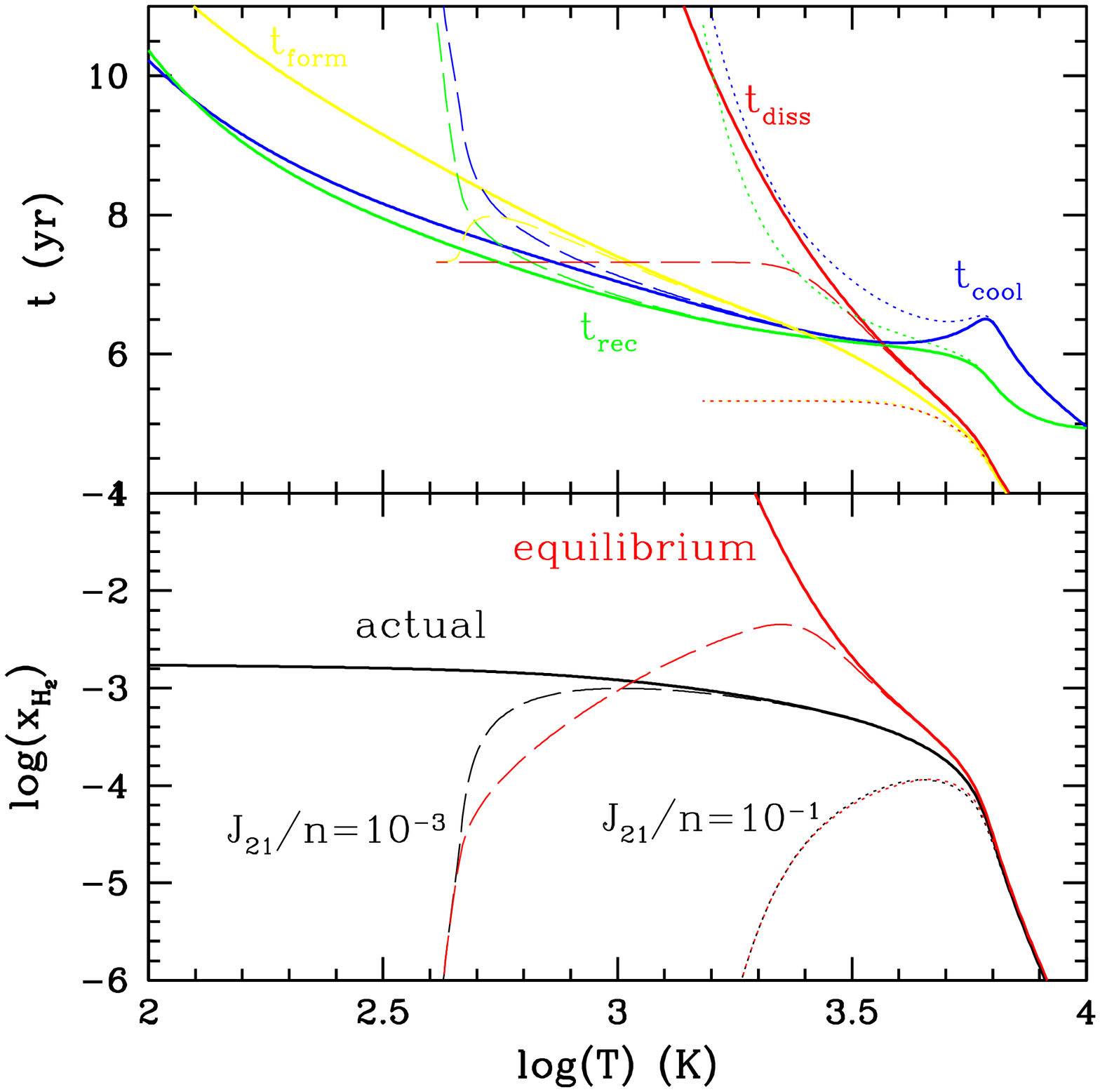}{3.3}{0.5}{-10}{-10}
\figcaption{\label{timescales}
The evolution of the ${\rm H_2}$ abundance as a function of
temperature, starting from $T> 10^{4}$K. Solid lines indicate the no
flux case; the dashed lines ($J_{21}/n=10^{-3}$) and
dotted lines ($J_{21}/n=10^{-1}$) illustrate the case when a
dissociating UV flux is present (see section
\S\ref{UV_external} for details; $J_{21}$ has units of $10^{-21}
{\rm erg \, s^{-1} \, cm^{-2} Hz^{-1} \, sr^{-1}}$, and $n$ has units
of ${\rm cm^{-3}}$). The bottom panel shows
the equilibrium (as given by equation (\ref{H2_equil}) and the actual
(computed by the full non-equilibrium chemistry code)) $\Htwo$ abundance, while the top panel shows the timescales for $\Htwo$
formation $t_{\rm form}$, dissociation $t_{\rm diss}$, gas cooling
$t_{\rm cool}$ and electron recombination $t_{\rm rec}$. Initially at high temperatures the ${\rm H_2}$ formation and destruction timescales are
short compared to $t_{\rm sys}={\rm min}(t_{\rm rec},t_{\rm cool})$,
and the ${\rm H_2}$ abundance is in equilibrium, $t_{\rm form}=t_{\rm
diss}$. Once $t_{\rm diss},t_{\rm form} > t_{\rm sys}$ at $T\sim
3700$K, the ${\rm H_2}$ abundance falls out of equilibrium, and
``freezes out'' at $x_{\Htwo} \sim 10^{-3}$.  When a dissociating UV
flux is present, $t_{\rm diss}$ asymptotes to a fixed value, instead of rising
exponentially. Eventually, $t_{\rm diss}$ becomes the shortest
timescale in the problem. At this point, the $\Htwo$ abundance falls
back into equilibrium, and decreases rapidly.}
\vspace{\baselineskip}

\subsection{Why is $x_{H_{2}} \sim 10^{-3}$?}

Here we show that the asymptotic abundance of $x_{H_{2}} \sim 10^{-3}$
may be {\it quantitatively} understood as a result of freeze--out processes.

The ${\rm H_{2}}$ will no longer evolve (``freeze-out'') if the ${\rm H_{2}}$
formation and dissociation timescales are much longer than any other
characteristic timescale in the system:
\begin{equation}
{\rm t_{form},t_{diss} \gg t_{rec},t_{cool}}
\end{equation}
where $t_{\rm form} \equiv
\frac{n_{\rm H_2}}{\dot{n}_{\rm form}} \approx
\frac{x_{\rm H_{2}}}{x_{\rm e} n k_{9}(T)}$ is the $\Htwo$ formation time, $t_{\rm diss} \equiv \frac{n_{\rm
H_2}}{\dot{n}_{\rm diss}} \approx \frac{1}{k_{17} (T)n x_{\rm e}}$ is
the $\Htwo$ dissociation time, $t_{\rm cool}$ is the cooling time, and
$t_{\rm rec} \equiv \frac{1}{x_{e}n k_{4}}$ is the electron recombination time. The cooling and
recombination timescales regulate the temperature $T$ and the
ionization fraction $x_{e}$, which in turn affect the $\Htwo$
abundance. 

We plot these timescales, as extracted from the full
non-equilibrium calculation, in the top panel of Fig.
(\ref{timescales}). As the gas cools, the dissociation timescale
$t_{\rm diss}$ grows rapidly and the gas falls out of
equilibrium. Shortly thereafter, $t_{\rm form}$ exceeds $t_{\rm
rec},t_{\rm cool}$, and the $\Htwo$ abundance freezes out.

The rapid growth of $t_{\rm diss}$ is easy to understand. The rates for collisional dissociation $k_{15}$ and charge exchange
$k_{17}$ ($\Htwo$ dissociation is primarily due to the latter)
decrease exponentially as the temperature drops. On other hand, all other
timescales $t_{\rm form}, t_{\rm cool}, t_{\rm rec}$ have only a
power-law temperature dependence. Therefore as the gas cools the
dissociation time rises exponentially and rapidly becomes much longer
than all other timescales.

Let us denote the temperature at which $t_{\rm diss} > t_{\rm rec},
t_{\rm cool}$ as $T_{\rm freeze}$. Since $t_{\rm diss}$ depends much
more sensitively on temperature than $t_{\rm rec},t_{\rm cool}$, it
exceeds them in very close succession. We can simply solve $t_{\rm diss}= t_{\rm rec}$ to obtain:
\begin{equation}
k_{17}(T)=k_{4}(T) \ \Rightarrow \ T_{\rm freeze}=3700 {\rm K}
\label{tfreeze}
\end{equation}
Up to this point, the $\Htwo$ abundance is in equilibrium. The
equilibrium value of $\Htwo$ at $T_{\rm freeze}$ is:
\begin{equation}
x_{\rm
H_{2}}^{\rm freeze} \approx k_{9}(T_{\rm freeze})/k_{17}(T_{\rm
freeze}) \approx 2 \times 10^{-3}
\label{xasymp}
\end{equation}
which is in excellent agreement with the calculated value, $\sim 1-2
\times 10^{-3}$.  

For $T<T_{freeze}$, $t_{\rm diss} \gg t_{\rm form}$, due to
exponential temperature dependence of $t_{\rm diss}$ discussed above. Thus, $\Htwo$ formation is more rapid
than $\Htwo$ destruction and $x_{\Htwo} \ge x_{\Htwo}^{\rm
asymp}$. However, if $t_{\rm form} > t_{\rm rec},t_{\rm cool}$ for $T
< T_{\rm freeze}$, then the $\Htwo$ abundance will freeze out. Since
$t_{\rm form} \approx t_{\rm diss} \approx t_{\rm rec}$ at $T_{\rm freeze}$, we can write:
\begin{equation}
\frac{t_{\rm
form}}{t_{\rm rec}} = \left( \frac{x_{\Htwo}}{x_{\Htwo}^{\rm asymp}}
\right) \left(\frac{T}{T_{\rm freeze}} \right)^{-1.7}
\end{equation}
Since $x_{\Htwo} \ge x_{\Htwo}^{\rm asymp}$, we conclude that $t_{form} >
t_{rec}$, for $T < T_{\rm freeze}$. Similarly, we have:
\begin{equation}
\frac{t_{\rm form}}{t_{\rm cool}} \propto
x_{\Htwo} T^{2} x_{\rm e}^{-1} 
\end{equation} 
(where we have assumed
$\Lambda_{\Htwo} \propto T^{4}$, a good approximation for $T<
3000$K). As before, $x_{\Htwo} \ge x_{\Htwo}^{\rm asymp}$, so the
minimum value of $t_{\rm form}/t_{\rm diss}$ is given by $x_{\Htwo}=
x_{\Htwo}^{\rm asymp}$. If we evolve the rate equations at fixed $x_{\Htwo}=
x_{\Htwo}^{\rm asymp}$, the quantity $T^{2}/x_{\rm e}$ increases with
time. Thus, $t_{\rm form} > t_{\rm cool}$ for $T < T_{\rm freeze}$. 

Since $t_{\rm diss},t_{\rm form} \gg t_{\rm rec}, t_{\rm cool}$, the
conditions for freeze-out are satisfied. The $\Htwo$ follows its
equilibrium abundance until $T_{\rm freeze}=3700$K, whereupon it freezes
out. Our simple estimate from timescale arguments (\ref{tfreeze}),
(\ref{xasymp}) for $T_{\rm freeze}$ and $x_{\rm asymp}$ agree
quantitatively with the full non-equilibrium calculation shown in Fig
(\ref{timescales}). This gives us confidence that we understand the
dominant physical processes at work.  

\myputfigure{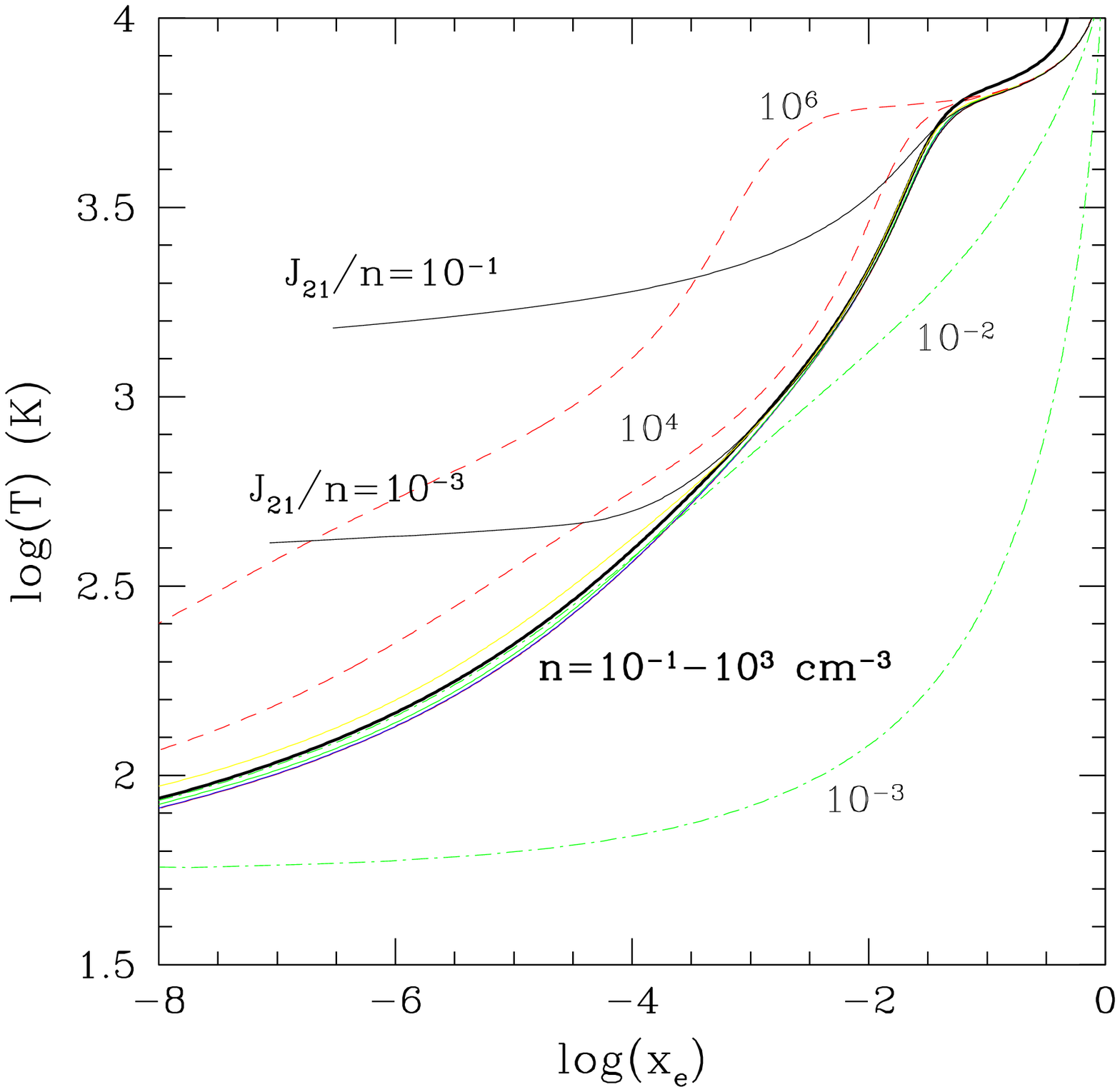}{3.3}{0.5}{-25}{-15}
\figcaption{\label{self_similar}
The evolution of the gas temperature against the ionization
fraction. The gas always cools along the same evolutionary pathway
(shown in bold), regardless of whether it cools isobarically or
isochorically, as long as the density stays within the limits $n
\approx 0.1-10^{3} \, {\rm cm^{-3}}$. Self-similarity is broken at low
and high density: at low densities (dot-dashed lines) cooling off the
CMB becomes important (the tracks here are for gas at z=20); at high
densities $n > 10^{3} {\rm cm^{-3}}$ (dashed lines), the ${\rm H_2}$
coolant reaches LTE and cooling becomes less efficient. Also shown are
the tracks when a dissociating UV flux is present (see section
\ref{feedback}), labelled by the value of $J_{21}/n$. The gas follows the universal track until $\Htwo$ is
dissociated and then recombines at constant temperature.}
\vspace{\baselineskip}

\subsection{Self-similarity}

All of the relevant timescales in this problem scale as $\propto
\frac{1}{n}$. Thus, their ratio is independent of density, and the evolution 
of the gas should be 'self-similar'. There are
three important variables controlling the evolution of the system:
$(T,x_{\rm e},n)$. Since the ratio of the timescales controlling
$(T,x_{\rm e})$ are independent of $n$, the evolution of the gas
in the $(x_{\rm e},T)$ plane should be independent of $n$ (note that
self-similarity would {\it not} be observed if $x_{\rm e}$ or $T$ were
plotted against $tn$, as in e.g. Shapiro \& Kang (1987), since in
this case there is one 'hidden variable' which is missing from the
plots).

In Figure \ref{self_similar}, we plot $T$ against $x_{\rm e}$ as
obtained by running the full chemistry code and find that indeed the
evolution of the gas is self-similar. This is subject to the following
caveats: (i) at low density and high redshift, Compton cooling of the
gas becomes effective; i.e., the gas cools faster by Compton cooling
than it recombines. Compton cooling is important in the regime $t_{\rm
cool}^{\rm Compton} < t_{\rm rec}$, i.e. when
\begin{equation}
n < 8 \times 10^{-3} \temphi^{0.7} \left( \frac{1+z}{20} \right)^{4}
\, {\rm cm^{-3}}
\label{Compton}
\end{equation}
Since the Compton cooling timescale is independent of density, the
self-similar scaling is broken in the low density regime, and the gas
cooling rate is enhanced relative to the recombination rate (see the
dot--dashed curves in Figure~\ref{self_similar}). (ii) At high
densities $n > 10^{4} {\rm cm^{-3}}$, the ${\rm H_2}$ cooling rate
$\propto n$ instead of $\propto n^{2}$, as the $\Htwo$ levels reach
LTE (Galli \& Palla 1998). Thus, the cooling time becomes independent
of density, rather than $t_{\rm cool} \propto \frac{1}{n}$, again
breaking the self-similar scaling. In this case, the recombination
rate is enhanced relative to the gas cooling rate (see dashed curves
in Figure \ref{self_similar}). However, between these two bounds the
track in $(x_{\rm e},T)$ space followed by the cooling gas (solid
lines in Figure \ref{self_similar}) is independent of the density and
indeed whether the gas cools isochorically or isobarically.

This convenient fact allows for considerable ease when estimating
timescales, since it collapses a system of two variables into one
variable, and allows one to 'evolve' a system without integrating the
full rate equations. For the convenience of the reader we provide a
fitting formula to this track:
\begin{eqnarray*}
{\rm log_{10}}(T)&=&10^{a_{o}+a_{1}x+a_{2} x^{2} +a_{3} x^{3} +
a_{4} x^{4}} ;\ \ x_{\rm e} > 10^{-2} \\ 
{\rm log_{10}}(T)&=&b_{o}+b_{1}x+b_{2}
x^{2} +b_{3} x^{3} + b_{4} x^{4} ; \ \ x_{\rm e} \le 10^{-2} 
\end{eqnarray*}
where $x\equiv {\rm log_{10}}(x_{\rm e})$ and $a_{0}=0.683,
a_{1}=0.379, a_{2}=0.522, a_{3}=0.299, a_{4}= 5.67 \times 10^{-2}$;
and $b_{0}=4.56, b_{1}=0.789,b_{2}=9.73 \times 10^{-2}, b_{3}=6.74
\times 10^{-3}$, and $b_{4}= 2.21 \times 10^{-4}$. This fit is valid
to within $\sim 0.5\%$ in dex over the range of scales shown in Figure
\ref{self_similar}.

The cooling path followed by the gas is also independent of the
initial temperature of the gas, as long as the initial ionization
fraction is somewhat greater than its value at freeze-out, $x_{\rm e}
\approx 2.4 \times 10^{-2}$, which translates into $T > 1.2 \times
10^{4}$K. This is because the gas recombines at nearly constant
temperature $T \sim 9000$K, and thus loses memory of its initial
temperature and ionization fraction (see the plateau in temperature in
Fig. \ref{full_chem}).

We can use the self-similar scaling between temperature and ionization fraction to estimate the timescale on which the ${\rm H_2}$ abundance reaches its
peak value. While the temperature as a function of time is difficult
to estimate, the ionization fraction as a function of time is
straightforward. It is given by:
\begin{equation}
x_{\rm e}(t) \approx \frac{x_{o}}{1+t/t_{\rm rec,o}}
\label{electron_fraction}
\end{equation}
where $x_{o} = 1$ is the initial electron abundance and $t_{\rm
rec,o}=1/k_{4}x_{o}n_{\rm H} =5.6 \times 10^{4} \density^{-1}
\, {\rm yr}$ (at 8000 K). At the freeze--out temperature $T_{\rm freeze} =
3700$K, the ionization fraction is $x_{\rm e} \approx 2.4 \times
10^{-2}$. Thus, ${\rm H_2}$ builds up to its
peak abundance on an approximate timescale $t \approx 2 \times 10^{6}
\density^{-1} \, {\rm yr}$.

For runaway collapse and fragmentation, we require $t_{\rm cool} <
t_{\rm dyn}$. For $x_{\rm H_2} \sim 10^{-3}$, we have:
\begin{equation}
\frac{t_{\rm cool}}{t_{\rm dyn}} = 0.22 \density^{-1/2} \left( \frac{T}{10^{3} {\rm K}  } \right)^{-3}
\end{equation} 
where we have assumed $\Lambda_{\rm H_2} \propto T^{4}$, a good
approximation for $T < 3000$K. In $T_{\rm vir} > 10^{4}$K halos, gas
will always contract via atomic cooling to high densities such that
$t_{cool} < t_{dyn}$. This is in contrast to $T_{\rm vir} < 10^{4}$K
halos, where only $\Htwo$ cooling is available, and only the
high-density gas at the center of the halo can fragment. 

In summary, we find that the universal abundance
$x_{\Htwo}\sim10^{-3}$ is the result of a ``freeze--out'' process. The
$\Htwo$ abundance follows its equilibrium value until the freeze--out
temperature $T_{\rm freeze} \approx 3700$K, at which point the $\Htwo$
formation and destruction timescales exceed the timescales on which
the system cools and recombines. At this temperature, the $\Htwo$ abundance freezes
out and stops evolving. In the absence of dissociating flux, $\Htwo$ formation and destruction timescales, as
well as gas cooling and recombination timescales, all depend on collisional
processes and scale as $\propto 1/n$. Hence, their ratio is independent of
density and the behavior of the gas is self--similar.

\section{Feedback and self-regulating star formation}
\label{feedback}

In the previous section, we have seen that in the absence of feedback,
$\Htwo$ always forms with an abundance of $x_{\Htwo}^{\rm asymp} \sim10^{-3}$. In this
section, we consider effects that can, in principle, modify this
conclusion. There are three principal feedback effects that can
potentially inhibit gas cooling and star formation in a disk: (i) UV
radiation and supernovae explosions heating the disk back up to $\sim
10^{4}$K, rendering the disk Toomre stable once again (ii) external
dissociating UV radiation preventing $\Htwo$ formation and cooling
(iii) internal dissociating UV radiation. In addition, we consider
(iv) Ly$\alpha$ photon trapping, which may cause the gas to become
radiation pressure dominated and prevent contraction to high
densities. We discuss each of these effects in turn.

\subsection{Gas heating and photoionization}

Our assumption of an isothermal disk is obviously an idealization;
both thermal instability and feedback from star formation implies that
the gas should develop a multi-phase structure. Nonetheless, as we
have seen in \S\ref{disk}, for low mass disks to remain Toomre
unstable, the dominant thermal phase must be cold (T$\sim 200$K). It
might be a concern that photoionizing radiation from forming stars
could heat the disk back up to $T \sim 10^{4}$K, stabilizing the disk
once again. The production rate of ionizing photons necessary for most
of the disk to be photoionized is:
\begin{eqnarray}
\nonumber
\dot{N}_{\rm ion} &\approx& \alpha_{\rm B} \int n_{\rm e}^{2} dV  \\
\nonumber
&=& 10^{56} \left( \frac{f_{\rm d}}{0.5} \right)^{3} \left( \frac{T_{\rm
gas}}{10^{4}
\, {\rm K}} \right)^{-1.7} \left( \frac{T_{\rm vir}}{5 \times 10^{4} \, {\rm
K}} \right)^{5/2} \\  &&\left( \frac{\lambda}{0.05} \right)^{-4} \left(
\frac{1+z}{10} \right)^{1.5} \ {\rm photons \, s^{-1}}  
\end{eqnarray} 
For a zero-metallicity Salpeter IMF with $0.1 < M_{*}/M_{\odot} <100$
when the lifetime of the starburst is $\tau \sim 10^{6}$yrs,
the stellar ionizing photon production rate is $\dot{N}_{\rm ion}
\approx 10^{53} {\rm \left( \frac{M_{*,{\rm tot}}}{10^{6} {\rm M_\odot}}
\right) \, photons \, \, s^{-1}}$, where $M_{*,{\rm tot}}$ is the total
mass in stars (Tumlinson \& Shull 2000). The entire disk gas mass
would be converted into stars before this stabilizing mechanism
becomes important. If the stellar IMF consists wholly of supermassive stars $M > 100 M_{\odot}$,
the ionizing photon production efficiency is an order of magnitude
greater (Bromm et al 2001c), and this feedback mechanism may be
marginally important. A likely much more
important feedback mechanism could be supernova explosions, which can
heat the gas to higher temperatures, and unbind it from the disk. A
detailed quantitative understanding of supernovae feedback is still
elusive (for recent progress see Efstathiou 2000, Mac-Low \& Ferrara
1999). We merely note that the effect of supernova explosions is
devastating for low mass halos, but it becomes progressively less
severe for $T > 10^{4}$K halos. The potential wells are deeper, the
gas contracts to the center of halos to yield higher binding energies,
and SNe are surrounded by a denser ambient medium, which can radiate
away the energy of the explosion more efficiently.

Another potential feedback process is production of secondary
electrons by X-ray sources. 
This could be an important effect if the first light sources were
mini--quasars (Haiman, Abel \& Rees 2000), or high-redshift supernovae
were an important source of X-rays (Oh 2001). This delays $\Htwo$
formation and cooling, since the secondary electrons produced increase
the rate of $\Htwo$ destruction by charge exchange processes (Kang \& Shapiro 1992). As
the gas gradually cools and recombines, the electron fraction falls
and $\Htwo$ production can proceed, unless the rate of X-ray heating
can balance gas cooling at $\sim 10^{4}$K to achieve stable thermal
equilibrium.  

\subsection{Photodissociation by an external UV field} 
\label{UV_external}

Here we consider the effects of an external photodissociating UV
radiation field. We show that $H_{2}$ formation and cooling depends
primarily on a single parameter, $J_{21}/n$, which controls the peak
$\Htwo$ abundance and thus the minimum temperature $T_{\rm min}$ a
parcel of gas can cool to. The value of $J_{21}/n$ is significantly
smaller in $T_{\rm vir} > 10^{4}$K halos: atomic cooling allows the
gas to contract to much higher initial densities. Thus, the effects of
an external UV radiation field are much less important, even if the
gas does not self-shield. Throughout the rest of this paper
$J_{21}/n$ is implicitly given in units of $10^{-21} {\rm erg \,
s^{-1} \, cm \, Hz^{-1} \, sr^{-1}}$. 

\subsubsection{The importance of $J_{21}/n$}

The timescale for photodissociation is given by (Draine \& Bertoldi 1996):
\begin{equation}
t_{\rm diss}=\frac{1}{k_{\rm diss}}= 2.1 \times 10^{4} J_{21}^{-1}
f_{\rm shield}^{-1} \, {\rm yr}
\end{equation}
where $J_{21}$ is the average flux in the LW bands in units of
$10^{-21} {\rm erg \, s^{-1} \, cm^{-2} Hz^{-1} \, sr^{-1}}$, and the factor
$f_{\rm shield}$ takes into account the effects of ${\rm H_2}$ self-shielding for a static clump of gas. For now we simply set $f_{\rm
shield}=1$; we will consider the effects of self-shielding
later, at the end of this section. Unlike all the other timescales
previously considered, $t_{\rm diss}$ is independent of
density. Photodissociation thus breaks the self-similarity of the
no-flux case, where all timescales scaled as $1/n$. Since
$t_{i}/t_{\rm diss} \propto J_{21}/n$ (where $i={\rm rec,cool,form}$),
the cooling history of the gas should depend only on the parameter
$J_{21}/n$. We show this in detail below. 

{\bf Dependence of $x_{\Htwo}^{\rm peak}$ on $J_{21}/n$} As the gas cools and
recombines from above $10^{4}$K, it initially follows the no-flux
case, where dissociation by charge exchange is the shortest timescale in the problem. However, as the temperature drops,
photodissociation will take over. From $t_{\rm diss}^{\rm
photo}=t_{\rm diss}^{\rm charge-exhange}$, this takes place when:
\begin{equation}
\left( \frac{J_{21}}{n} \right) \approx 1.5 \times 10^{-2} {\rm exp}\left[ - \frac{3700 \, {\rm
K}}{{\rm T}} \right] \left( \frac{x_{\rm e}}{2.4 \times 10^{-2}} \right)
\label{flux_condition}
\end{equation}
At late times (low $T$ and $x_{e}$), LW dissociation will always
dominate ${\rm H_2}$ destruction. Instead of rising exponentially as the temperature drops,
the ${\rm H_2}$ dissociation time remains constant (provided
self-shielding is unimportant), becoming the shortest timescale in the
problem as $t_{\rm cool}, t_{\rm rec}$ rise. Thus, the ${\rm H_2}$
abundance rises to a peak value before steadily declining once $t >
t_{\rm diss}$.

The peak abundance of ${\rm H_2}$ can be estimated as follows. The
cooling gas follows the no flux evolution along the universal track
until the condition in equation (\ref{flux_condition}) is satisfied at
some branch-out temperature $T_{\rm bo}$. Observe that equation
(\ref{flux_condition}) is a function of only the temperature, since
$x_{\rm e}(T)$ is given by the fit to the universal cooling
track. There are two possibilities: (1) $T_{\rm bo} < T_{\rm freeze}$,
i.e. $J_{21}/n < 1.5 \times 10^{-2}$. In this case, $x_{\Htwo}^{\rm peak}
\sim x_{\Htwo}^{\rm asymp} \sim 10^{-3}$. The $\Htwo$ abundance
reaches the asymptotic freeze-out value and falls out of equilibrium, until $t_{\rm rec},t_{\rm cool}$
increase sufficiently that $t_{\rm diss} < t_{\rm rec},t_{\rm cool}$.
At this point, ${\rm H_2}$ falls back into equilibrium and its abundance
steadily decreases. (2) $T_{\rm bo} > T_{\rm freeze}$, i.e., $J_{21}/n > 1.5
\times 10^{-2}$. In this case, photodissociation becomes important
before freeze-out can occur. ${\rm H_2}$ is always in equilibrium, and
we can compute the peak abundance simply from the equilibrium abundance of
${\rm H_2}$ at $T_{\rm bo}$, i.e. $x_{\rm H_{2}}(x_{\rm e}(T_{\rm
bo}),T_{\rm bo})$.

These expectations are clearly borne out in runs with the full
non-equilibrium chemistry code. In Figure (\ref{timescales}), we plot the various timescales discussed
above and the $\Htwo$ abundance as a function of temperature. We
consider two different flux levels: $J_{21}/n = 10^{-1}$ (dotted lines) and $J_{21}/n = 10^{-3}$ (dashed
lines), which bracket the critical value of $J_{21}/n = 1.5 \times
10^{-2}$.For $J_{21}/n = 10^{-1}$, the ${\rm H_2}$ abundance never reaches freeze
out but always follows the equilibrium value, whereas for $J_{21}/n =
10^{-3}$, the gas reaches the freeze--out value of $x_{\rm H_{2}} \sim
10^{-3}$ and falls out of equilibrium, only to fall back into
equilibrium when $t_{\rm diss} = {\rm min}(t_{\rm rec},t_{\rm
cool})$. The $\Htwo$ abundance declines rapidly thereafter. In Figure
\ref{self_similar}, we see that the cooling gas follows the universal
track until an abrupt transition takes place, corresponding to $t_{\rm
diss} = {\rm min}(t_{\rm rec},t_{\rm cool})$, at which point the gas
stops cooling and recombines at constant temperature.

{\bf Dependence of $T_{\rm min}$ on $J_{21}/n$} The results above
suggest that the minimum temperature $T_{\rm min}$ the gas can cool to
depends on $J_{21}/n$. We have verified this by running the full
chemistry code for a parcel of gas at different densities and under different flux levels, assuming isobaric
cooling (the result is independent of the time the gas has available
to cool $t_{\rm H}$ as long as $t_{\rm diss} < t_{\rm H}$. If $t_{\rm
diss} > t_{\rm H}$, then photodissociation is in any case unimportant).
In Figure \ref{Jn_plot}, we plot the minimum temperature a parcel of
gas cools down to, $T_{\rm min}$ against $J_{21}/n$. Where the cooling
of the gas is in the self-similar regime $n \sim 10^{-1} - 10^{4} {\rm
cm^{-3}}$, the curves lie on top of one another regardless of the
absolute values of $J_{21},n$, confirming that $T_{\rm min}$ is a
function only of one variable $J_{21}/n$. When $n > 10^{4} {\rm
cm^{-3}}$ then as before this scaling behavior is broken: the cooling
time becomes independent of density and the gas cools less
efficiently. In particular $T_{\rm min}$ is no longer a function of
$J_{21}/n$ but just $J_{21}$ (see dashed line). We explore this regime
in more detail in \S\ref{UV_internal}.

We can compute $T_{\rm min}$ analytically. Since the gas
follows the universal track until $t_{\rm diss}(J_{21}) < {\rm min}(t_{\rm rec},t_{\rm
cool})$, we can solve the equation:
\begin{equation}
t_{\rm diss}={\rm min}(t_{\rm rec},t_{\rm cool})
\end{equation}
to obtain the temperature $T_{\rm min}$ at which the $\Htwo$ is
dissociated and the gas stops cooling (this argument only holds for
$J_{21}/n < 1.5 \times 10^{-2}$.  For $J_{21}/n > 1.5 \times 10^{-2}$
then $t_{\rm diss} < t_{\rm cool}$ always holds. The gas can
nonetheless cool by a small amount, which must be calculated by solving the energy equation). We
plot the result as points in Figure \ref{Jn_plot}. The simple estimate
$t_{\rm diss}=t_{\rm cool}$ shows remarkably good agreement with the
results of the full chemistry code.

\subsubsection{The cooled gas fraction in halos}

The significance of the previous result is as follows. In order for the gas to
cool down to a temperature $T_{\rm min}$, it must have $\left(
\frac{J_{21}}{n} \right) < \left( \frac{J_{21}}{n} \right)_{\rm crit}$; in
particular, in the presence of a given dissociating flux $J_{21}$, the
gas must have $n > n_{\rm crit}$. In halos with $T_{\rm vir} <
10^{4}$K, ${\rm H_2}$ is the only available coolant; if initially $n <
n_{\rm crit}$ the gas will never cool to $T_{\rm min}$. For this
reason, $\Htwo$ formation and cooling is thought to be
inhibited in $T< 10^{4}$K halos due to the effects of an external UV
dissociating background (Haiman, Rees \& Loeb 1997, Ciardi et
al. 2000, Haiman, Abel \& Rees 2000, Machacek et al. 2001). However,
the effects of an external UV background are much less severe for $T >
10^{4}$K halos, primarily because atomic cooling contracts the gas to
sufficiently high densities that the $\Htwo$ formation and cooling
timescales are shorter than the photodissociation time. Initially,
when the gas density is low, $\Htwo$ formation will be suppressed, and
the gas will cool by atomic cooling, contracting until $n > n_{\rm
crit}$ and $\Htwo$ cooling takes over. From Figure \ref{Jn_plot}, we
see that in order to cool the gas down to $\sim 500$K, we require $n >
n_{\rm crit} \sim 10^{3} J_{21} {\rm cm^{-3}}$.

>From equations (\ref{disc_density}) and (\ref{central_density}) we see
that almost all of the baryons in the disk will satisfy this criterion
(as the gas cools it will be compressed further, increasing the gas
fraction which satisfies $n > n_{\rm crit}$). By contrast, in $T_{\rm
vir} < 10^{4}$K halos, where there is no means of initially compressing
the gas, only a small fraction of the gas exists at $n > n_{\rm
crit}$. We can quantify this statement by assuming that the gas in
$T_{\rm vir} < 10^{4}$K halos is initially isothermal at the virial
temperature of the halo. The density profile of the gas is then given by demanding
hydrostatic equilibrium in an NFW halo (Makino et al. 1998). We adopt
this profile to compute the fraction of baryonic mass of a halo above a
given number density $M_{\rm b}(> n)$. Halos which collapse at similar
redshift are roughly self-similar, and have a fixed fraction of their
mass above a given density, independent of their mass. 

In Figure \ref{frac_cool}, we compute the fraction of gas which can
cool down to a temperature $T$ for halos at $z=15$, as a function of
the external UV radiation field $J_{21}$. For comparison, the
radiation field corresponding to $n_{\gamma}$ ionizing photons per
baryon in the universe is $J_{21} \approx \frac{h_{P} c}{ 4 \pi}
n_{\gamma} n_{b} (1+z)^{3} \times 10^{21} \approx 10 n_{\gamma}
\left(\frac{1+z}{16}\right)^{3}$ (where $n_{b}$ is the comoving baryon
number density, and $h_{P}$ is the Planck constant). Only an extremely
small fraction of the gas can cool to low temperatures $T < 500$K required for a
reasonably small Jeans mass, $M_{J}=10^{4} (T/500~{\rm K})^{3/2}
(n/10^{4}~{\rm cm^{-3}})^{-1/2} \, \msolar$.

\myputfigure{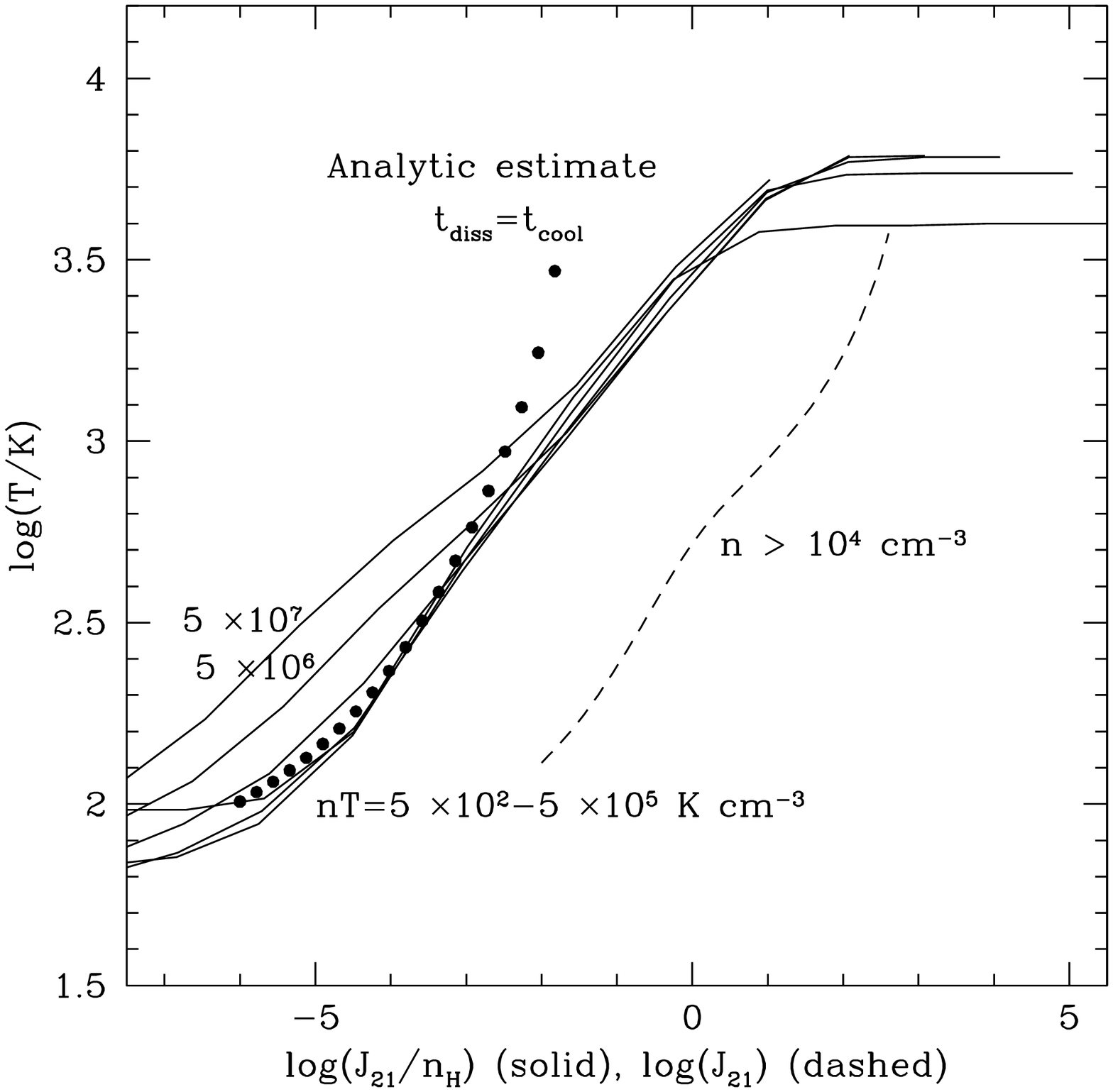}{3.3}{0.5}{-10}{-10}
\figcaption{\label{Jn_plot}
The minimum temperature $T_{\rm min}$ a gas parcel can cool down to,
as a function of $J_{21}/n$, for $nT=5 \times
10^{2},10^{3},10^{4},10^{5},10^{6},10^{7} \, {\rm K \,
cm^{-3}}$ (solid curves). These results were obtained by integrating the rate
equations for an isobarically cooling parcel of gas. The cooling
depends only on a single variable $J_{21}/n (\propto t_{\rm
cool}/t_{\rm diss})$; this can be seen from
the fact that almost all the curves lie on top of one
another. This scaling behaviour is only broken when the gas approaches
high densities $n> 10^{4} {\rm cm^{-3}}$ (i.e., the $nT= 5 \times
10^{6}, 5 \times 10^{7} \, {\rm K \,
cm^{-3}}$ curves) at which point the cooling
time becomes independent of density and $t_{\rm cool}/t_{\rm diss}
\propto J_{21}$. In this case, the minimum temperature the gas can
cool down is not given by $J_{21}/n$ but by $J_{21}$ alone (dashed line). The points represent
the analytic estimate of $T_{\rm min}$ as a function of $J/n$ by setting $t_{\rm
diss}=t_{\rm cool}$; in the range $J/n < 1.5 \times 10^{-2}$ where
this estimate is valid, the agreement with the full calculation is
remarkably good.}
\vspace{\baselineskip}

A similar quantity has been computed by Machacek et al. (2001) in
numerical simulations of halo formation and gas cooling. In their
Figure 3, they plot the fraction of gas $f_{\rm gas}$ that has cooled to
$T < 0.5 T_{\rm vir}$ and $\rho > \rho_{\rm threshold}$ as a function of
$T_{\rm vir}$ and find a strong correlation between $f_{\rm gas}$ and
$T_{\rm vir}$. The reason for this is that the cooling function of ${\rm H_2}$
depends strongly on $T$. Since the cooling function $\Lambda_{\Htwo}
\propto T^{4}$ is much greater at higher $T$, $n_{\rm crit}$ is
correspondingly lower for high $T_{\rm vir}$ halos and thus $f_{\rm
gas}$ is higher. We find reasonable agreement (to within a factor of two in the cold mass fraction), if we plot our
results in this fashion. Here we choose to compute the fraction which
cools to a fixed temperature $T$, since ultimately it is the final
temperature which determines the Jeans mass. By our criterion, there
is no correlation of $f_{\rm gas}$ with $T_{\rm vir}$. By comparison,
virtually all of the disk gas in halos with $T_{\rm vir} > 10^{4}$K halos
can cool to low temperatures $T< 500$K, due to the high densities in
the disk. Overall, while levels of flux comparable to that required
for the universe to be fully reionized will strongly suppress cooling
in $T_{\rm vir} < 10^{4}$K halos, they have negligible impact on ${\rm
H_2}$ formation and cooling in $T_{\rm vir} > 10^{4}$K halos.

\myputfigure{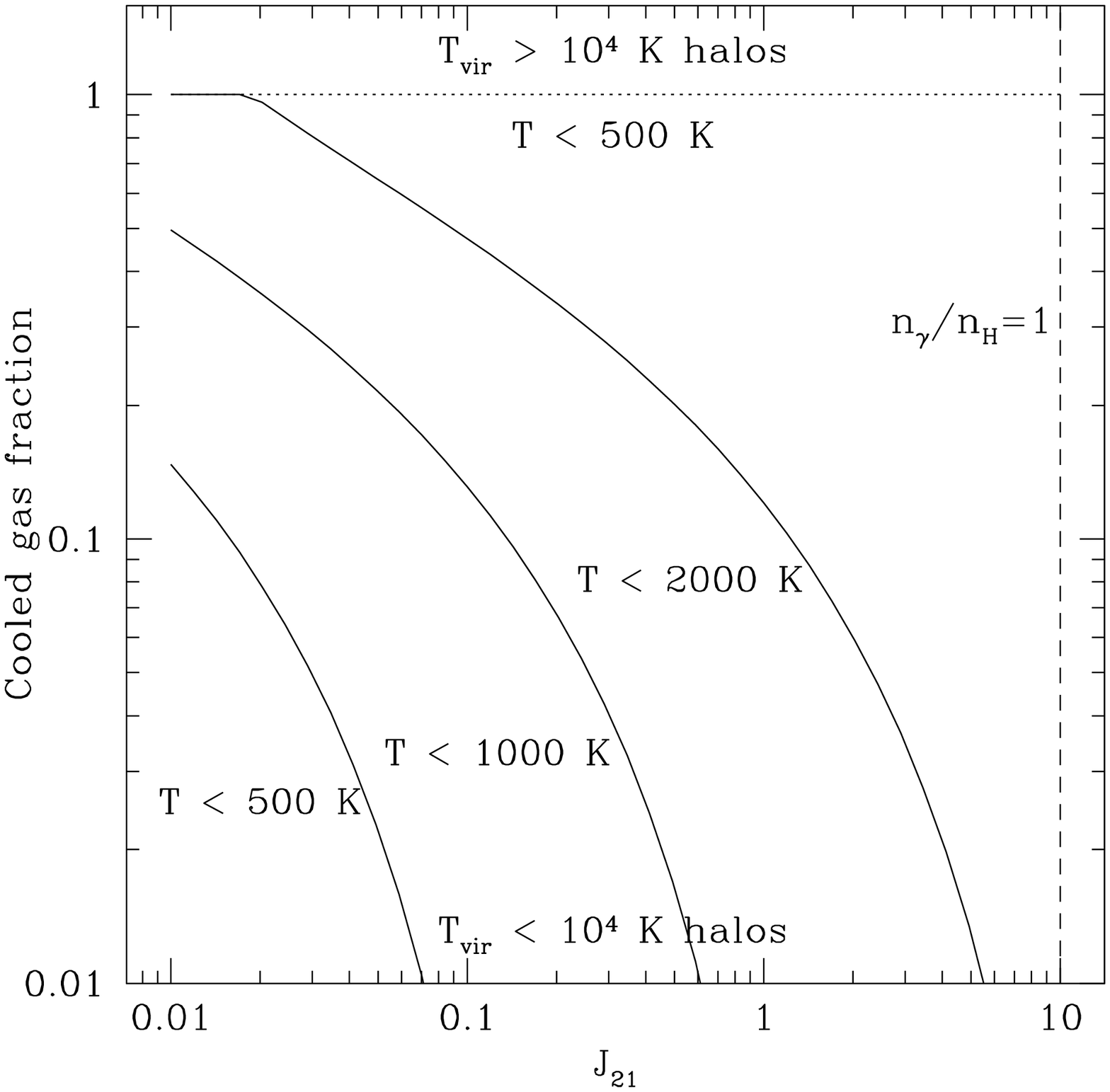}{3.3}{0.5}{-25}{-10}
\figcaption{\label{frac_cool}
The fraction of gas in a halo that can cool below a given temperature as a
function of the external UV radiation field, for halos at z=15. The
radiation field corresponding to 1 ionizing photon per baryon in the
universe is marked. In halos with $T_{\rm vir} < 10^{4}$K (solid lines), only a
small fraction of the gas is at sufficiently high density to be
unaffected by external radiation fields. By contrast, in halos where
atomic cooling operates, the gas contracts to such high densities that
virtually all of the disk gas can form ${\rm H_2}$ and cool to low
temperatures.}
\vspace{\baselineskip}

\subsubsection{Self-shielding}

We now consider the role of self-shielding. We have previously seen
that gas in disks can
form ${\rm H_2}$ and cool for all reasonable values of the external UV
radiation field even if we neglect self-shielding. We therefore confine
ourselves to a few general remarks. Self-shielding is much
more important in $T_{\rm vir} > 10^{4}$K halos than in lower mass
halos: after the gas has settled in the disk, column densities are
higher by a factor $\lambda^{-2} \sim 400$. If ${\rm H_2}$ forms with
the asymptotic abundance of $x_{\rm H_{2}} \sim 10^{-3}$, then from
equation (\ref{column_density}), the ${\rm H_2}$ column density is:
$N_{\rm H_2}(r) \approx 10^{20} {\rm exp}(-r/2 R_{\rm d}) \left(
\frac{f_{\rm d}}{0.5} \right) \left(\frac{T_{\rm vir}}{5 \times 10^{4}\, 
{\rm K}} \right)^{1/2} \left(\frac{\lambda}{0.05} \right)^{-2} \left( 
\frac{1+z}{10} \right)^{3/2}\ {\rm cm^{-2}}$. 

There are two extremes cases to be considered. If the gas is static
$v/b \ll 1$, then the LW flux is attenuated by a factor: $f_{sh}= {\rm
min}\left[ 1, \left (\frac{N_{\rm H_{2}}} {10^{14} {\rm cm^{-2}}}
\right)^{-0.75} \right] \ll 1$ (Draine \& Bertoldi 1996; this fitting formula is accurate to within a factor of 2 for $N_{\rm H_{2}} < 5 \times 10^{20}{\rm cm^{-2}}$, at which point the LW flux is attenuated by 5 orders of magnitude). In this regime, self-shielding is extremely strong and the
gas is impervious to external UV radiation. However, if there are
significant velocity gradients with $v/b \gg 1$, then the gas remains
optically thin to LW radiation until the damping wings of the LW lines
overlap, at $N_{\rm H_{2}} \approx 10^{22} {\rm cm^{-2}}$ (Draine \&
Bertoldi 1996, Glover \& Brand 2000; in particular, see Fig. 2 and 3
of Glover \& Brand 2000, where the fraction of radiation in the LW
bands absorbed is computed as a function of ${\rm H_2}$ column
density. For $N_{\rm H_{2}} \sim 10^{20} {\rm cm^{-2}}$, only $\sim 10
\%$ of the radiation is absorbed). The magnitude of self-shielding
therefore depends on the velocity field of the gas, which probably
lies closer to the $v/b \gg 1$ regime, particularly as the disk cools
by $\Htwo$ cooling to temperatures $T_{\rm gas} \ll T_{\rm vir}$, and
thus $b \ll V_{\rm
circ}$. We have seen that gas
in the disk is unaffected by photodissociating UV radiation even if
self-shielding is unimportant. On the other hand, the ${\rm H_2}$
optical depth affects the efficiency of ${\rm H_2}$ dissociation by
internal sources; if the optical depth is high, then a large fraction
of the energy emitted in the LW bands by stars goes towards
dissociating ${\rm H_2}$. We now turn to $\Htwo$ photodissociation by
internal UV sources.

\subsection{Photodissociation by internal UV radiation}
\label{UV_internal}

$\Htwo$ photodissociation by internal sources is likely to be the
dominant source of feedback. Following Glover \& Brand (2000), we can
estimate the dissociation efficiency as follows. A single 100${\rm
M_{\odot}}$ metal free star produces $\dot{N}_{\rm dis}\sim 10^{49} \
{\rm photons \ s^{-1}}$ in the 11.15-13.6 eV range. Let us assume that
a fraction $f_{\rm abs}\sim 0.1$ of these photons are absorbed in the
disk (as is appropriate if $N_{\rm H_2} \sim 10^{20} {\rm cm^{-2}}$;
see Figures 2 \& 3 in Glover \& Brand 2000) and $f_{\rm dis} \sim 0.2$
of excitations lead to dissociations. Then over its main sequence
lifetime $t_{\rm ms} \sim 3 \times 10^{6}$yr, the star will dissociate
$M_{\rm gas}
\approx
f_{\rm dis} f_{\rm abs} m_{\Htwo} \dot{N}_{\rm dis} t_{\rm ms}/x_{\Htwo} \sim
10^{7} {\rm M_\odot}$ of gas of all its $\Htwo$, assuming $x_{\Htwo}
\sim 10^{-3}$ (see also Omukai \& Nishi (1999)).  This could imply a
star formation efficiency as low as $f_{\rm star} \sim 10^{-5}$.

The efficiency with which cooling and star formation can still proceed
depends on the amount of clumping and fragmentation that can take
place before internal sources of radiation turn on. Dense clumps have
short free-free times and may be able to collapse before all the
$\Htwo$ in them is dissociated. While the $\Htwo$ in the center of a
cooling clump may initially be shielded from dissociating radiation
(since $N_{\Htwo} \gg 10^{14} {\rm cm^{-2}}$ for most clumps),
ultimately if $t_{\rm diss}\equiv M_{\Htwo}/\dot{M}_{\Htwo} < t_{\rm
dyn}$, where $M_{\Htwo}$ is the mass of $\Htwo$ in the clump and
$\dot{M}_{\Htwo}$ is the $\Htwo$ dissociation rate, then the clump
cannot collapse.

For a uniform density clump, the requirement that $t_{\rm dyn} <
t_{\rm diss}$ translates into the requirement that the clump lie at a
distance $D > 10\left( \frac{\dot{N}_{\rm dis}}{10^{49} s^{-1}}
\right)^{1/2} \denfour^{-7/12} \left( \frac{M_{\rm clump}}{10^{4} {\rm
M_\odot}} \right)^{-1/6} \left(\frac{f_{\rm abs}}{10^{-2}}
\right)^{-1/2}$pc from the nearest dissociating source. By comparison,
the virial radius is $r_{\rm vir} \approx 2 \left( \frac{T_{\rm
vir}}{5 \times 10^{4} \, K} \right)^{1/2} \redshift^{-3/2}$kpc, and
the disk scale length is $R_{\rm d}\approx \frac{\lambda}{\surd 2}
r_{\rm vir} \approx 70$pc. Even if $\Htwo$ is completely dissociated
in the disk, it may still be possible for dense clumps to form in the
halo (where the radiation field from stars is weaker), either
fragmenting and forming stars in the halo itself or falling into the
disk and accreting gas there.

There is a hard limit to the temperature a parcel of gas exposed to
dissociating radiation can cool to. This is because at high densities
$n> 10^{4} {\rm cm^{-3}}$, collisional excitation and de--excitation
dominate, the gas falls into LTE, and the cooling time becomes
independent of density. We can find this minimum temperature by
setting $t_{\rm cool}(x_{\Htwo},T)=t_{\rm diss}(J_{21})$ for the
universal $\Htwo$ fraction $x_{\Htwo} \sim 10^{-3}$. The result is
shown in Figure \ref{Jn_plot} (dashed curve). If the dissociating flux exceeds a
critical value of $J > J_{\rm crit} \approx 10^{3}$, then cooling
below $\sim 4000$K is impossible, independent of the density of the
gas. Such flux levels are easily attainable from internal stellar
radiation fields: a $100 \, {\rm M_\odot}$ metal free star would
generate a radiation field $J_{21} \approx 10^{3} \left(
\frac{\dot{N}_{\rm dis}}{10^{49} s^{-1}} \right) \left(\frac{D}{20 
{\rm pc}}\right)^{-2}$ in the LW bands. Since $D \sim 20$pc is of
order the size of typical cooling clumps, it implies that once a
single star forms within a clump, subsequent cooling and fragmentation
within the clump will be suppressed. 

If internal photodissociation is a very efficient source of feedback,
then star formation initiated by $\Htwo$ cooling cannot take place in
the disk, and is likely to proceed only in the halo (where the sources
of radiation are spread farther apart). If the disk has sufficiently
low spin to be Toomre unstable at $\sim 10^{4}$K, then unstable clumps
could form in the disk, cooling isothermally by atomic cooling. As
long as they can cool to extremely high densities ($n> 10^{12} {\rm
cm^{-3}}$), then stellar--mass size clumps could still form. Omukai
(2001) has suggested that the gas cools initially by Ly$\alpha$ and two photon
emission for $n < 10^{7} {\rm cm^{-3}}$, and then by free-bound
emission from ${\rm H^{-}}$ until $n \sim 10^{16} {\rm cm^{-2}}$, when
the gas becomes optically thick to continuum radiation. This cooling
route does not permit the formation of $\Htwo$ by three--body
reactions at high densities when cooling from high initial
temperatures. Unless sufficient $\Htwo$ can form initially to cool the
gas down to $T < 2000$K, collisional dissociation completely
overwhelms the three--body $\Htwo$ formation rate (see rates by Galli,
Palla \& Salpeter 1982). We have found the equilibrium values of
$\Htwo$ for $ T > 5000$K to be negligibly small, even for extremely high densities $n \gg 10^{8} \, {\rm cm^{-3}}$. In principle, clumps
cooling via atomic cooling could cool to sufficiently high densities
so as to reach sub--solar fragmentation masses (Omukai 2001). This
leads to a somewhat counterintuitive possibility: within 
photodissociation regions where $J_{21} > 10^{3}$ and $\Htwo$
formation and cooling is suppressed, the fragmentation masses could
be {\it smaller} than in regions where $\Htwo$ formation and cooling
can can take efficiently.

A possible obstacle to contraction to such high densities is radiation
pressure from Ly$\alpha$ photon scattering; we examine this issue in
the next subsection. If the majority of disks are Toomre stable and
photodissociation by internal UV sources is very efficient, then
efficient star formation may have to await not just $T > 10^{4}$K
halos, but halos capable of self-enriching themselves in metals, with
sufficiently deep potential wells that metals ejected by supernovae
rain back down on the disk in a galactic fountain, rather than being
lost to the IGM. From local observations, the critical circular
velocity appears to be $v \sim 130 {\rm km \, s^{-1}}$ (Martin
1999). For metal-free stars radiative mass loss is likely to be 
inefficient (Kudritzki 2000), and except for stars which collapse
directly to black holes and do not eject metals, $\sim$half the stellar mass is
converted to metals (Heger \& Woosley 2001). Thus, the fraction of gas
processed into stars roughly corresponds to the metallicity enrichment, $f_{\rm star} \sim Z$. Studies of gas cooling (Hellsten
\& Lin 1997, Bromm et al. 2001b) suggest that $Z \sim 10^{-3} {\rm
Z_\odot} \sim 10^{-5}$ is the critical metallicity at which metal line cooling
allows the gas to cool down to $ T < 100$K, implying that a prior star
formation efficiency of $f_{\rm star} > 10^{-5}$ is required with
${\rm H_2}$ cooling before metal line cooling can take over. This 
seems compatible (given the many uncertainties) with our estimate of
$f_{\rm star} \sim 10^{-5}$.

\subsection{Ly$\alpha$ photon trapping and radiation pressure} 

Collapse and fragmentation can be halted if the opacity of the contracting gas
rises sufficiently that the energy density of trapped radiation
becomes high and radiation pressure support becomes important. The
condition for this to occur is:
\begin{equation}
L t_{\rm trap} >  E_{\rm bind} \approx \frac{G M^{2}}{R}
\label{rad_pressure}
\end{equation}
where $L$ is the luminosity of the radiation source, $t_{\rm trap}$ is
the timescale on which photons are trapped within the collapsing
cloud, and $E_{\rm bind}$ is the binding energy of the cloud. If the
only source of radiation is cooling radiation and the system is in
free fall collapse, then $L \approx E_{\rm bind}/t_{\rm dyn}$ and
condition (\ref{rad_pressure}) translates into the condition $t_{\rm
trap} > t_{\rm dyn}$. Rees \& Ostriker (1977) have emphasized that if
the only radiation source is cooling, then even if all radiation is
trapped, radiation pressure cannot halt an overall collapse, since the
resulting energy density is only half that required to support the
cloud ($P_{\rm gas}=2/3 U_{\rm gas}$, but $P_{\rm rad}=1/3 U_{\rm
rad}$; in addition, in our case $\sim$ half of the energy is radiated
in the two photon continuum, rather than in the Ly$\alpha$ line, and
can escape more easily). Nonetheless, they note that radiation
pressure could become sufficiently important so as to boost the Jeans
mass and halt fragmentation. We now investigate this possibility, and
also consider the case when an internal radiation source provides
additional photons to provide pressure support.

When the system is isothermally contracting by atomic cooling at $\sim
8000$K and the gas is largely neutral, the dominant opacity
source is Ly$\alpha$ scattering. The optical depth at line center to
escaping Ly$\alpha$ photons across a contracting clump of mass $M_{J}
\approx 3 \times 10^{5} \tempeight^{3/2} \denfour^{-1/2} {\rm
M_{\odot}}$ is $\tau = 1.2 \times 10^{10} \denfour \left( \frac{r}{\rm 7 \, pc}
\right)$, which is enormous. However, due to the finite line width the
photons diffuse in both space and frequency, escaping from the cloud
when they have scattered sufficiently far from line center. This
problem has been investigated by a number of authors (Adams 1975,
Neufeld 1990), who find:
\begin{eqnarray}
\nonumber
t_{\rm trap} = 15  \, t_{\rm cross} \ \ 10^{3} \le \tau \le 10^{6} \\ 
 t_{\rm trap} = 15  \, t_{\rm cross} \left( \frac{\tau}{10^{6}} \right)^{1/3}
 T_{4}^{1/6} \ \ \tau \ge  10^{6}  
\label{t_trap}
\end{eqnarray}
where $t_{\rm cross}=R/c$ is the light crossing time. In addition,
Bonilha et al. (1979) find from Monte Carlo calculations that dust and
velocity gradients reduce the trapping time by a factor $f_{\rm dust}
= \frac{1}{(1 +0.9 \delta)^{0.97}}$ and $f_{\rm v}=
\frac{1}{1+0.027 \eta^{1.5}}$ respectively, where
$\delta=\frac{t_{\rm trap}}{t_{\rm cross}} \tau_{\rm dust}$ and $\eta
\approx 1.5 \left(\frac{\Delta V}{b} \right)$, where $\Delta V$ is the
velocity gradient across the cloud and $b$ is the Doppler
parameter. Since we only consider gas of almost zero metallicity ($Z<
10^{-3} {\rm Z_\odot}$) we ignore the effects of dust; in addition,
since we have $\Delta V\sim b$ and thus $f_{\rm v} \sim 1$, velocity
gradients are unimportant. From equation (\ref{t_trap}) we find that
at $n \sim 10^{4} {\rm cm^{-3}}$ then $t_{\rm trap} \approx 350 t_{\rm cross} \approx 8 \times
10^{-3} t_{\rm dyn}$, and radiation pressure is unimportant. If the
clump does not fragment and $n r^{3} = {\rm const}$, then $t_{\rm
trap}/t_{\rm dyn} \propto n^{8/18} \propto M^{-8/9}$ and thus at $n
\sim 5 \times 10^{8} {\rm cm^{-3}}$, when $M_{J} \sim 10^{3} \, {\rm
M_\odot}$, radiation pressure could potentially become comparable to
the thermal gas pressure. In particular, radiation pressure could
significantly impede the contraction of a massive clump of gas.

In reality, the gas is likely to fragment as it cools. If we only
consider the photon diffusion time across a Jeans length, then for
isothermal cooling $M_{J} \propto n R_{J}^{3} \propto n^{-1/2}$, i.e.
$R_{J} \propto n^{-1/2}$. This yields $t_{\rm trap}/t_{\rm dyn}
\propto n^{1/6} \propto M^{-1/3}$. In this case, the Jeans mass can
fall far below a solar mass before radiation pressure ever becomes
important.

We now consider the case where an internal ionizing source, such as
stars or QSOs provide additional Ly$\alpha$ photons from their HII
regions, and consider whether collapse and fragmentation in the
(largely neutral) gas beyond the surrounding regions can be halted by
radiation pressure. Consider a cooling clump of mass $M_{J}$ at
$T=10^{4}$K. In order for $P_{\rm rad} = \frac{1}{3}{U_{\rm rad}}
\approx \frac{1}{3} \frac{L_{\rm Ly\alpha} t_{\rm trap}}{V} > P_{\rm therm} = nk_{\rm b}T$, we require:
\begin{equation} 
\dot{N}_{\rm Ly\alpha} \approx \dot{N}_{\rm ion} > 4 \times 10^{50} \left( \frac{M_{J}}{10^{6} {\rm
{\rm M_\odot}}} \right)^{1/3} {\rm photons \ s^{-1}}
\end{equation}
where we assume that each ionizing photon results in the production of
a Ly$\alpha$ photon. By comparison, for metal free stars with
$M_{*} > 100 {\rm M_\odot}$, we have $\dot{N}_{\rm ion} \sim 10^{50} \left(
\frac{M_{*}}{100 \, {\rm M_\odot}} \right) \ {\rm photons \ s^{-1}}$
(Tumlinson \& Shull 2000, Bromm et al. 2001c); therefore, only $\sim
10^{-3}$ of the mass of a $10^{6} {\rm M_\odot}$ clump need fragment into
such stars for radiation pressure to become important. However, this
is still unlikely to completely prevent smaller subclumps from collapsing. The critical star formation efficiency required to maintain
$P_{\rm rad}=P_{\rm gas}$ as a function of the Jeans mass is
\begin{equation}
f_{\rm star} \equiv \frac{\rho_{*}}{\rho_{\rm g}} \approx 10^{-3} \left( \frac{M_{J}}{10^{6} {\rm
M_{\odot}}} \right)^{-2/3}.
\label{Ly_SF}
\end{equation}
Thus, when $M_{J} \sim 300 \, {\rm M_\odot}$, $f_{\rm star} \sim
100\%$ of the gas must fragment into stars to halt collapse. We
conclude that Ly$\alpha$ photon pressure from stars can halt the
overall contraction of a cooling cloud, but it does not maintain the
Jeans mass above a certain mass scale. Smaller subclumps have shorter
trapping times and require unattainably high internal Ly$\alpha$
photon production rates to halt collapse. However, Ly$\alpha$ pressure
is likely to act as a feedback mechanism regulating the efficiency
with which such clumps collapse, since it prevents the ambient medium
from cooling and condensing to high densities.

In particular, radiation pressure from stars or AGN can significantly retard or halt the free fall collapse of the gas; if cooling gas becomes radiation pressure
supported, then entropy fluctuations
will be smoothed out and further fragmentation could be very
inefficient. This could well imply star formation efficiencies of order that given by equation (\ref{Ly_SF}). Ly$\alpha$ photon pressure could regulate the accretion of gas onto a dense clump, as well as increase the scale height of the
disk. It is therefore likely to be an important source of feedback
until supernova explosions become dominant. A full quantitative
investigation of the effects of Ly$\alpha$ photon trapping in the
early universe is beyond the scope of this paper (previous authors
have studied the effects of radiation pressure in related
contexts; see, e.g., Cox 1985; Bithell 1990; Haehnelt 1995).

\section{Conclusions}
\label{conclusions}

In this paper, we have examined the cooling of gas in the first halos
at high redshift with $T_{\rm vir} > 10^{4}$K, which are able to cool
via atomic line cooling. Previous generations of halos with $T_{\rm
vir} < 10^{4}$K, in which the only available coolant is $\Htwo$
molecules, are expected to have their cooling strongly suppressed by
feedback processes (Haiman, Abel \& Rees 2000; Omukai \& Nishi
1999). Thus, the gas in these larger halos is likely to still be of
nearly primordial composition.  Our main findings can be summarized as
follows:

\begin{enumerate}
\item The gas should cool isothermally initially and settle toward a
rotationally supported disk at the center of the dark matter
halo. Unless the halo has unusually low angular momentum or a large
fraction ($f_{\rm d} > 0.5$) of the gas cools to form the disk (by
contrast, present day dwarfs typically only have $f_{\rm d} \sim
0.3$), the majority of such disks will be gravitationally stable by
the Toomre criterion. An additional coolant is therefore required to
lower the gas sound speed and promote gravitational instability. We
identify this coolant as $\Htwo$.

\item In gas which has been collisionally ionized and heated to 
a temperature of $T>10^4$K, $\Htwo$ molecules form with a universal
fraction $x_{\Htwo} \sim 10^{-3}$, independent of initial density or
temperature for a wide range of initial conditions. This universal
fraction is not an equilibrium value. Rather, it can be understood as
the result of a 'freeze--out'. The $\Htwo$ formation and destruction
timescale become long compared to all other timescales at a
temperature $T_{\rm freeze} \sim 3700$K; $x_{\Htwo}
\sim 10^{-3}$ thus reflects the equilibrium value of $\Htwo$ at $T_{\rm
freeze}$, the temperature at which the $\Htwo$ abundance stops
evolving. Independence of density can be understood from the fact that
all collisional timescales scale as $\propto 1/n$, and thus the ratio
of timescales is independent of density; in particular, the gas
follows a universal track in the $(x_{e}, T)$ plane. Independence of the initial temperature can be understood from
the fact that the gas recombines isothermally at $\sim 9000$K, losing
memory of the initial conditions. 

\item We have examined the feedback from UV radiation fields and conclude
that $\Htwo$ photodissociation from an external UV background,
comparable to that required to reionize the universe, is unlikely to
be important. This is because atomic cooling contracts the gas to such
high initial densities that the $\Htwo$ formation and cooling
timescales are shorter than the UV photodissociation timescales. This
is in contrast to $T< 10^{4}$K halos where only a small fraction of
$\Htwo$ at the center of the halo is at sufficiently high density to
promote cooling within a photodissociation time; the fraction of gas in such halos which cools is much
lower. We argue that radiation pressure due to Ly$\alpha$ photon
trapping could conceivably limit the efficiency
of star formation. This is because only a small fraction of the gas
needs to be converted into stars for a collapsing clump to become
radiation pressure supported. Another possible source of pressure
support, which we have not examined, arises from turbulent velocity
fields (Padoan 1995). 
\end{enumerate}

While the fact that ${\rm H_2}$ can form out of metal-free gas cooling
from $T> 10^{4}$K and cool the gas to $\sim 10^{2}$ K is a robust
result, we stress that our picture of how this enables gas to settle
into cold, gravitationally unstable disks at the center of halos is
considerably more uncertain. At stake is not only a large variation in
the efficiency of star formation in low mass halos and thus the
redshift of reionization (see Figures \ref{frac_spiral} and
\ref{redshift_evol}), but also the mass scale at which clumps form,
which bears upon the stellar IMF and perhaps upon the question of
supermassive black hole formation. In particular, the degree of
fragmentation of the gas is highly uncertain. These important
questions can only be addressed with some degree of confidence by
high-resolution numerical simulations which are able to track the
detailed gas hydrodynamics, chemistry and cooling, paralleling the
pioneering work already done for $T_{\rm vir} < 10^{4}$K halos (Abel
et al. 2000, Bromm et al.  2001a). The main hope of this paper is to
stimulate further work along these lines.

\acknowledgements

We thank D. Neufeld and T. Abel for useful comments; Volker Bromm,
Jordi Miralda-Escud\'e, Andrey Kravtsov, and Martin Rees for
stimulating conversations; and Joop Schaye for correcting a factor of 2 error. SPO was supported by NSF grant AST-0096023. ZH was supported by NASA
through the Hubble Fellowship grant HF-01119.01-99A, awarded by the
Space Telescope Science Institute, which is operated by the Association of
Universities for Research in Astronomy, Inc., for NASA under contract
NAS 5-26555.

\markright{APPENDIX}
\centerline{APPENDIX: REACTION RATES AND CROSS SECTIONS}
\markright{APPENDIX}

\begin{center}
\begin{tabbing}
\hspace{1.1cm}\= Reaction \= \hspace{0.5cm}\= \= \hspace{2.8cm} \= Rate Coefficient (${\rm cm^3}$ ${\rm sec^{-1}})$ \hspace{6.5cm}\= Reference \\
\rule[-0.05in]{7.1in}{0.01in}\\
(1)  \> $H$\hspace{0.6cm}\=$+ \hspace{0.2cm} e^-  $\>$ \rightarrow H^+$\hspace{0.4cm}\=$+\hspace{0.2cm}2e^-  $ \> $5.85\times 10^{-11} T^{\frac{1}{2}}{\rm exp} (-157809.1/T)(1+T_5^{\frac{1}{2}})^{-1}       $\hspace{4.7cm} \=     [1] \\
(2)  \> $He     $\>$+\hspace{0.2cm}  e^-  $\>$ \rightarrow He^+   $\>$+\hspace{0.2cm}2e^-  $ \> $2.38\times 10^{-11} T^{\frac{1}{2}}{\rm exp}
(-285335.4/T)(1+T_5^{\frac{1}{2}})^{-1}   $                         \> [2] \\
(3)  \> $He^+   $\>$+\hspace{0.2cm}  e^-  $\>$ \rightarrow He^{++} $\>$+\hspace{0.2cm}2e^-  $ \> $5.68\times 10^{-12} T^{\frac{1}{2}}{\rm exp} (-631515.0/T)
         (1+T_5^{\frac{1}{2}})^{-1}$                                \> [1] \\
(4)  \> $H^+    $\>$+\hspace{0.2cm}  e^-  $\>$ \rightarrow H      $\>$+\hspace{0.2cm} h\nu $ \> $8.40\times 10^{-11} T^{-\frac{1}{2}} T_3^{-0.2}
             (1+T_6^{0.7})^{-1}    $                                \> [1] \\
(5)  \> $He^+   $\>$+\hspace{0.2cm}  e^-  $\>$ \rightarrow He     $\>$+\hspace{0.2cm} h\nu $ \>  {\rm see expression in reference}                   \> [1] \\
(6)  \> $He^{++}$\>$+\hspace{0.2cm}  e^-  $\>$ \rightarrow He^+   $\>$+\hspace{0.2cm} h\nu $ \> $3.36\times 10^{-10} T^{-\frac{1}{2}} T_3^{-0.2}
             (1+T_6^{0.7})^{-1}     $                               \> [1] \\

(7)  \> $H      $\>$+\hspace{0.2cm}  H^+  $\>$ \rightarrow H_2^+  $\>$+\hspace{0.2cm} h\nu $ \>  {\rm see expression in reference}                  \> [3] \\
(8)  \> $H_2^+  $\>$+\hspace{0.2cm}  H    $\>$ \rightarrow H_2    $\>$+\hspace{0.2cm} H^+  $ \> $6.40\times 10^{-10}                               $ \> [4] \\
(9*)  \> $H      $\>$+\hspace{0.2cm}  e^-  $\>$ \rightarrow H^-
$\>$+\hspace{0.2cm} h\nu $ \> $1.4\times 10^{-18} T^{0.928}\exp(-T/16200)$ \> [5] \\
(10) \> $H      $\>$+\hspace{0.2cm}  H^-  $\>$ \rightarrow H_2    $\>$+\hspace{0.2cm} e^-  $ \> $1.30\times 10^{-9}                                $ \> [6] \\
(11) \> $H_2^+  $\>$+\hspace{0.2cm}  e^-  $\>$ \rightarrow 2H      $\>$ \hspace{0.2cm}      $ \> $1.68\times 10^{-8} (T/300)^{-0.29}        $ \> [7]\\
(12) \> $H_2^+  $\>$+\hspace{0.2cm}  H^-  $\>$ \rightarrow H_2    $\>$+\hspace{0.2cm} H    $ \> $5.00\times 10^{-6} T^{-\frac{1}{2}}               $ \> [8] \\
(13*) \> $H^-    $\>$+\hspace{0.2cm}  H^+  $\>$ \rightarrow 2H     $\>$
\hspace{0.2cm}      $ \> $5.7\times10^{-6}T^{-\frac{1}{2}} + 6.3\times10^{-8} - 9.2\times10^{-11}T^{\frac{1}{2}} + 4.4\times10^{-13}T$ \> [5] \\
(14) \> $H_2    $\>$+\hspace{0.2cm}  e^- $\>$ \rightarrow H       $\>$+\hspace{0.2cm} H^-  $ \> $2.70\times 10^{-8} T^{-\frac{3}{2}}\exp(-43000/T) $ \> [9] \\
(15*) \> $H_2    $\>$+\hspace{0.2cm}  H   $\>$ \rightarrow 3H      $\>$
$ \> $1.067\times10^{-10}\left(\frac{T}{\rm eV}\right)^{2.012}
\exp[-4.463 (\frac{T}{\rm eV})^{-1}]/( 1+0.2472 (\frac{T}{\rm eV})^{3.512})  $  \> [10] \\
(16) \> $H_2    $\>$+\hspace{0.2cm}  H_2 $\>$ \rightarrow H_2    $\>$+\hspace{0.2cm}2H     $ \>  {\rm see expression in reference}                   \> [11] \\
(17*) \> $H_2    $\>$+\hspace{0.2cm}  H^+ $\>$ \rightarrow H_2^+
$\>$+\hspace{0.2cm} H     $ \>  $3\times 10^{-10}\exp(-21050/T)~(T_4<1); 1.5\times 10^{-10}\exp(-14000/T)~(T_4\geq1)$
\> [5] \\
(18) \> $H_2    $\>$+\hspace{0.2cm}  e^- $\>$ \rightarrow 2H     $\>$+\hspace{0.2cm}e^-    $ \>  $4.38\times 10^{-10}{\rm exp}(-102000/T)T^{0.35}$   \> [12] \\
(19) \> $H^-    $\>$+\hspace{0.2cm}  e^- $\>$ \rightarrow H      $\>$+\hspace{0.2cm}2e^-   $ \>  $4.00\times 10^{-12}T{\rm exp}(-8750/T)$            \> [12] \\
(20) \> $H^-    $\>$+\hspace{0.2cm}  H   $\>$ \rightarrow 2H     $\>$+\hspace{0.2cm}e^-    $ \>  $5.30\times 10^{-20}T^{2.17}{\rm exp}(-8750/T)$     \> [12] \\
(21) \> $H^-    $\>$+\hspace{0.2cm}  H^+ $\>$ \rightarrow H_2^+  $\>$+\hspace{0.2cm}e^-    $ \>  {\rm see expression in reference}                   \> [12] \\
\rule[0.05in]{7.1in}{0.01in}\\
\> \> \> \> \> Cross-section $({\rm cm^2})$ \> \\
\rule[0.05in]{7.1in}{0.01in}\\
(22) \> $H      $\>$+\hspace{0.2cm}h\nu  $\>$ \rightarrow H^+
$\>$+\hspace{0.2cm}e^-   $ \> $6.3\times10^{-18}(\frac{\rm
13.6eV}{h\nu})^4\exp(4-4(\tan^{-1})\epsilon/\epsilon)/[1-\exp (-2\pi/\epsilon)]
; \epsilon\equiv\sqrt{\frac{h\nu}{\rm 13.6}-1} $                    \> [13] \\
(23*) \> $He     $\>$+\hspace{0.2cm}h\nu  $\>$ \rightarrow He^+
$\>$+\hspace{0.2cm}e^-   $ \>  $0.694\times10^{-18}
[(\frac{h\nu}{\rm eV})^{1.82} + (\frac{h\nu}{\rm eV})^{3.23}]^{-1}$                                   \> [14] \\
(24) \> $He^+   $\>$+\hspace{0.2cm}h\nu  $\>$ \rightarrow
He^{++}$\>$+\hspace{0.2cm}e^-   $ \>  $1.575\times10^{-18}(\frac{\rm
54.4eV}{h\nu})^4\exp(4-4(\tan^{-1})\epsilon/\epsilon)/[1-\exp (-2\pi/\epsilon)]
; \epsilon\equiv\sqrt{\frac{h\nu}{54.4}-1} $                                    \> [13] \\
(25*) \> $H^-    $\>$+\hspace{0.2cm}h\nu  $\>$ \rightarrow H $\>$+\hspace{0.2cm}e^-   $ \>  $7.928\times10^5(\nu-\nu_T)^{\frac{3}{2}}\nu^{-3}${\rm~for~}$h\nu\geq h\nu_T=0.755~{\rm eV}$     \> [15] \\
(26*) \> $H_2^+  $\>$+\hspace{0.2cm}h\nu  $\>$ \rightarrow H $\>$+\hspace{0.2cm}H^+   $ \>     {\rm see expression in reference}  \> [12] \\
(27) \> $H_2    $\>$+\hspace{0.2cm}h\nu  $\>$ \rightarrow H_2^+  $\>$+\hspace{0.2cm}e^-   $ \>  {\rm see expression in reference}                    \> [12] \\
(28) \> $H_2    $\>$+\hspace{0.2cm}h\nu  $\>$ \rightarrow 2H     $\>$ \hspace{0.2cm}      
$ \>$\sigma(\nu)=3\times 10^{-18}${\rm~for~}$11.26~{\rm eV}<h\nu<13.6~{\rm eV}$\> [16]\\
\rule[0.05in]{7.1in}{0.01in}
\end{tabbing}
\end{center}

{\footnotesize 
1. Cen (1992); 
2. Black (1978); 
3. Rawlings et~al.~(1993); 
4. Karpas et~al.~(1979);
5. Galli and Palla (1998); 
6. de Jong (1972); 
7. Nakashima et~al.~(1987); 
8. Dalgarno \& Lepp (1987); 
9. Hirasawa (1969); 
10. Dove \& Mandy (1986);
11. Lepp \& Shull (1983); 
12. Shapiro \& Kang (1987); 
13. Osterbrock (1974);
14. Haardt \& Madau 1996; 
15. Abel et al. (1997);
16. Allison \& Dalgarno (1970) and Dalgarno \& Stephens (1970).}

\vspace{\baselineskip} 

The table above lists the chemical reactions included in our chemistry network.
For reference, we have marked (with a ``*'') the reactions whose adopted rates
are different from the compilation in Haiman, Rees \& Loeb (1996).

\end{document}